\documentclass[aps,twocolumn,prd,showpacs,showkeys,preprintnumbers,superscriptaddress,nobibnotes,floatfix,longbibliography,notitlepage,nofootinbib]{revtex4-1}

\def\apj{ApJ}%
\def\apjs{ApJS}%
\def\jcap{JCAP}%
\def\prd{Phys.\ Rev.\ D}%

\pdfoutput=1
\usepackage{amsmath}
\usepackage{amsfonts}
\usepackage{amssymb}
\usepackage{mathrsfs}
\usepackage{graphicx}
\usepackage{color}
\usepackage{bbding}
\usepackage{tabularx}
\usepackage[usenames,dvipsnames,table]{xcolor}
\usepackage[normalem]{ulem}
\usepackage{makecell}
\usepackage{xspace}
\usepackage[colorlinks=true,allcolors=blue]{hyperref}
\usepackage{multirow}
\usepackage{makecell}
\usepackage{upgreek}
\usepackage[capitalise]{cleveref}
\usepackage{listings}
\usepackage{fontawesome5}
\usepackage{placeins}

\newcommand{\cosmion}{\texttt{cosmion}\xspace}

\graphicspath{{figures/}{figures_appendix/}}

\begin{document}

\title{Simulation of thermal conduction by asymmetric dark matter\\ in realistic stars and planets}

\author{Hannah Banks}
\email{hmb61@cam.ac.uk}
 \affiliation{Department of Applied Mathematics and Theoretical Physics, University of Cambridge, Wilberforce Road, Cambridge, CB3 0WA, UK }
\author{Stephanie Beram}
\email{stephanie.beram@queensu.ca}
\affiliation{Department of Physics, Engineering Physics and Astronomy, Queen's University, Kingston ON K7L 3N6, Canada}
\affiliation{Arthur B. McDonald Canadian Astroparticle Physics Research Institute,  Kingston ON K7L 3N6, Canada}
\author{Rashaad Reid}
\email{19yrr@queensu.ca}
\affiliation{Department of Physics and Astronomy, University of Waterloo, Waterloo ON N2L 3G1, Canada}
\affiliation{Department of Physics, Engineering Physics and Astronomy, Queen's University, Kingston ON K7L 3N6, Canada}

\author{Aaron C. Vincent}
\email{aaron.vincent@queensu.ca}
\affiliation{Department of Physics, Engineering Physics and Astronomy, Queen's University, Kingston ON K7L 3N6, Canada}
\affiliation{Arthur B. McDonald Canadian Astroparticle Physics Research Institute,  Kingston ON K7L 3N6, Canada}
\affiliation{Perimeter Institute for Theoretical Physics, Waterloo ON N2L 2Y5, Canada}

\begin{abstract}
Dark matter captured in stars can act as an additional heat transport mechanism, modifying fusion rates and asteroseismoloigcal observables. Calculations of heat transport rates rely on approximate solutions to the Boltzmann equation, which have never been verified in realistic stars. Here, we simulate heat transport in the Sun, the Earth, and a brown dwarf model, using realistic radial temperature, density, composition and gravitational potential profiles. We show that the formalism developed in Ref.~\cite{Banks:2021sba} remains accurate across all celestial objects considered, across a wide range of kinematic regimes, for both spin-dependent and spin-independent interactions where scattering with multiple species becomes important. We further investigate evaporation rates of dark matter from the Sun, finding that previous calculations appear robust. Our Monte Carlo simulation software \texttt{cosmion} is publicly available. \href{https://github.com/aaronvincent/cosmion}{\faGithub}  
\end{abstract}

\maketitle

\section{Introduction}
\label{sec:intro}
Whilst a compelling portfolio of evidence for dark matter (DM) has been collated, its microphysical nature and properties continue to elude us.   With an extraordinarily broad landscape of possible candidates generating a similarly vast array of phenomenological consequences, the hunt to uncover the true nature of DM continues to be of upmost priority to the fundamental physics community. To date, experimental searches for particle DM have primarily focused on the GeV-TeV scale Weakly Interacting Massive Particle (WIMP), and other WIMP-adjacent models with similar masses and Weak-scale (or smaller) interaction cross sections with the Standard Model (SM). Earth-based direct detection experiments searching for elastic scattering between halo dark matter and heavy nuclei or electrons via such interactions in shielded underground laboratories have led the way, with limits on spin-independent (SI) DM-nucleus scattering now falling below $10^{-47.5}$ cm$^2$ for a DM mass $m_\chi \simeq 30$ GeV  and below $10^{-41}$ cm$^2$ for spin-dependent (SD) interactions \cite{LUX-ZEPLIN:2022qhg}. 

The possible existence of such DM-SM couplings generate a variety of other opportunities to probe DM. Indeed, they necessarily lead to the capture of DM particles in astrophysical objects including planets, stars and stellar remnants, following scattering to velocities below the local escape velocity set by the gravitational potential, $\phi(r)$. Once captured, continued interactions with SM nuclei enable the DM particles to thermalize, and, thanks to a mean free path $\ell_\chi$ that exceeds the typical thermal transport length scale, transport heat outwards. If present in sufficient quantities this can result in observable modifications to the internal structure and properties of the astrophysical object in question. Although the cross section for ``optimal'' heat transport lies well above direct detection limits, such stellar observables can provide a verification of these results that is independent of any systematics to do with the Earth and laboratory-based experiments. They may additionally become competitive if objects can be identified in environments with much higher DM densities, e.g. near the galactic center. 

Two formalisms have traditionally been used to compute the effects of DM-mediated heat transport  in stars and to a lesser extent, in planets. Both involve approximate solutions to the 7-dimensional integro-differential Boltzmann collision equation. These are the Spergel \& Press (henceforth SP) \cite{Spergel1985EffectInterior} method valid for long mean free paths, and the Gould \& Raffelt (GR) \cite{Gould1990} approximation in the opposing limit of Local Thermal Equilibrium (LTE). In a recent publication \cite{Banks:2021sba}, we showed via explicit Monte-Carlo integration, that the SP method produces a more reliable model of the energy transport profile over a large range of DM masses, interaction cross sections, and interaction types scaling with velocity and exchanged momentum as may be expected in e.g. generic non-relativistic effective operators \cite{Fan2010Non-relativisticDetection,Fitzpatrick2013TheDetection,Catena2015FormTheories}.
The major limitation of this work was that, whilst realistic stellar temperature and collisional target (nuclei) distributions were used, the gravitational potential was approximated as a simple harmonic oscillator (SHO), as in the original simulations performed by Gould \& Raffelt \cite{Gould1990CosmionLimit}. This had the major advantage in that the trajectories could be solved analytically at all times, leading to a major computational speedup. Although this is a good approximation when the DM particles are sufficiently massive ($m_\chi \gtrsim 5$ GeV) to be confined within the stellar core where the density is approximately constant, at lower DM masses and in stars with different density profiles, one would expect a departure from SHO behavior. Additionally, whereas both the GR and SP formalisms can be straightforwardly extended to spin-independent scattering where the dark matter interacts coherently with every nucleon, this has never been simulated nor verified in the context of heat transport in a realistic star. 

The aims of this work are therefore to extend the simulations presented in Ref. \cite{Banks:2021sba} to
1) Integrate particle trajectories in arbitrary potential wells allowing for the study of DM mediated heat transport in a wider variety of astrophysical bodies including the Sun, the  Earth, a brown dwarf and a red giant star, and
2) 
Include scattering from multiple different isotopes to investigate the accuracy of generalizing the SP and GR formalisms to spin-independent interactions.

We begin in Sec.~\ref{sec:review} by reviewing the physics governing  DM heat transport in astrophysical bodies and the existing, commonly adopted formalisms which constitute approximate solutions in specific limiting regimes. In Sec. ~\ref{sec:mc} we then outline the Monte-Carlo procedure that we use to solve for the behaviour of a given thermally conducting DM population numerically. In Sec.~\ref{sec:results} we examine, in turn, the heat transport resulting from both SD- and SI- interactions of DM with nuclei in a variety of different astrophysical bodies and compare our results to those predicted by the two conventional formalisms. We conclude in Sec. \ref{sec:conclusion}. Appendix \ref{app:kepler} provides some details on how we treat particles leaving the star, a quickstart guide to the \texttt{cosmion} code can be found in Appendix~\ref{app:quickstart} and Appendix~\ref{app:collisionrate} offers some further information on the collision rates per nuclear species as relevant to spin-independent interactions, and their treatment within the \texttt{cosmion} code.  

\begin{figure*}[ht!]
    \centering
    \includegraphics[width=\columnwidth]{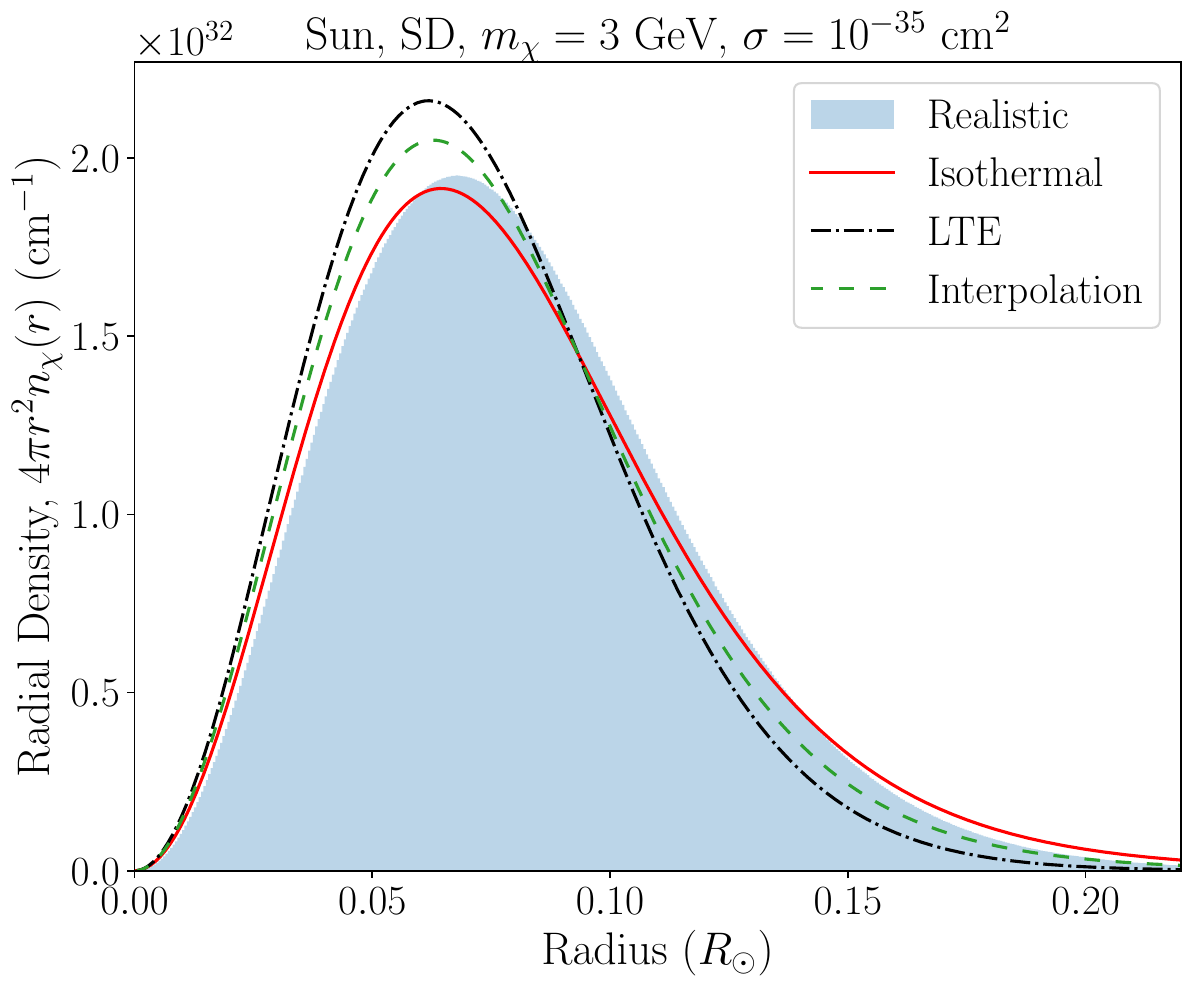}\includegraphics[width=\columnwidth]{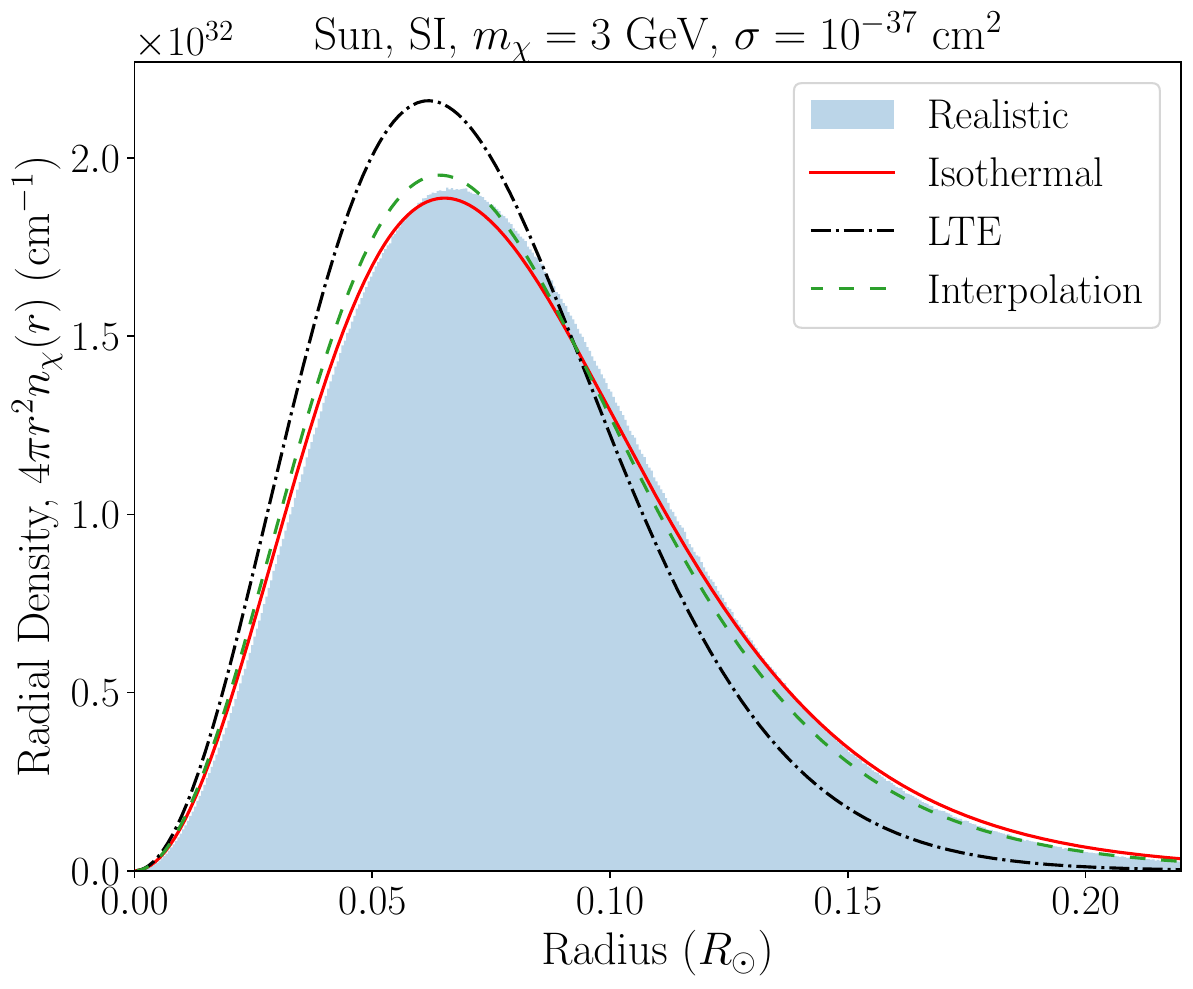}
    \caption{Radial distributions of $m_\chi=3$ GeV DM particles in the Sun with indicated constant scattering cross-sections $\sigma_0$, for spin-dependent (SD, left) and spin-independent (SI, right) interactions with nucleons. The shaded regions show the distributions obtained in our Monte Carlo simulations, while the lines are the analytically-predicted distributions. The isothermal line is the prediction made for the realistic potential by the Spergel \& Press formalism, and the LTE line is that made by the Gould \& Raffelt formalism. The dashed green line is an interpolation between the two, as proposed in Ref.~\cite{Scott2009}. }
    \label{fig:SunRadial}
\end{figure*}

\begin{figure*}[h]
    \centering
    \includegraphics[width=\columnwidth]{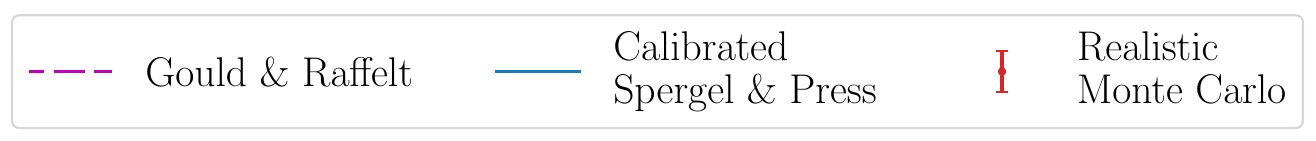}
    \includegraphics[width=\columnwidth]{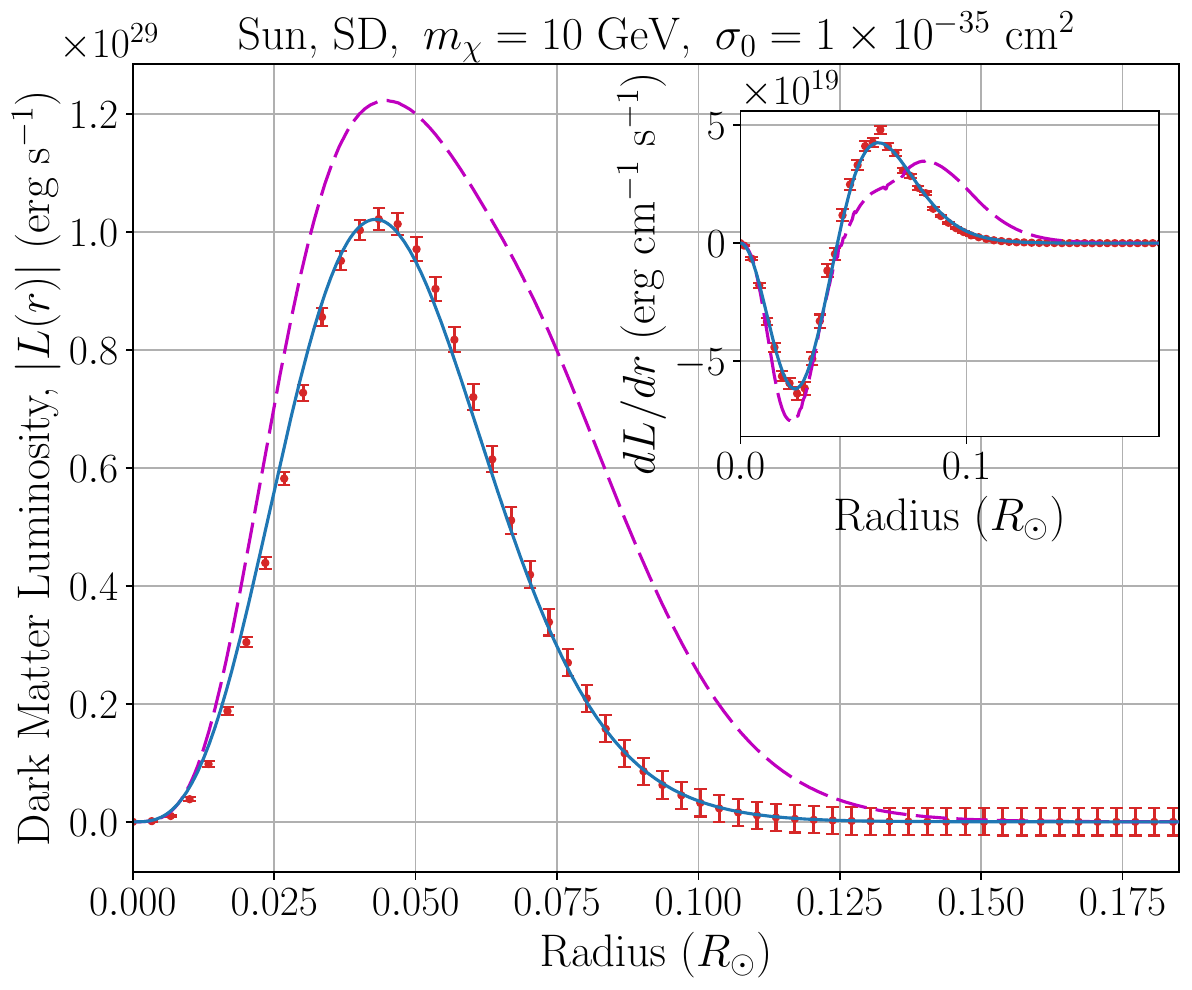}\includegraphics[width=\columnwidth]{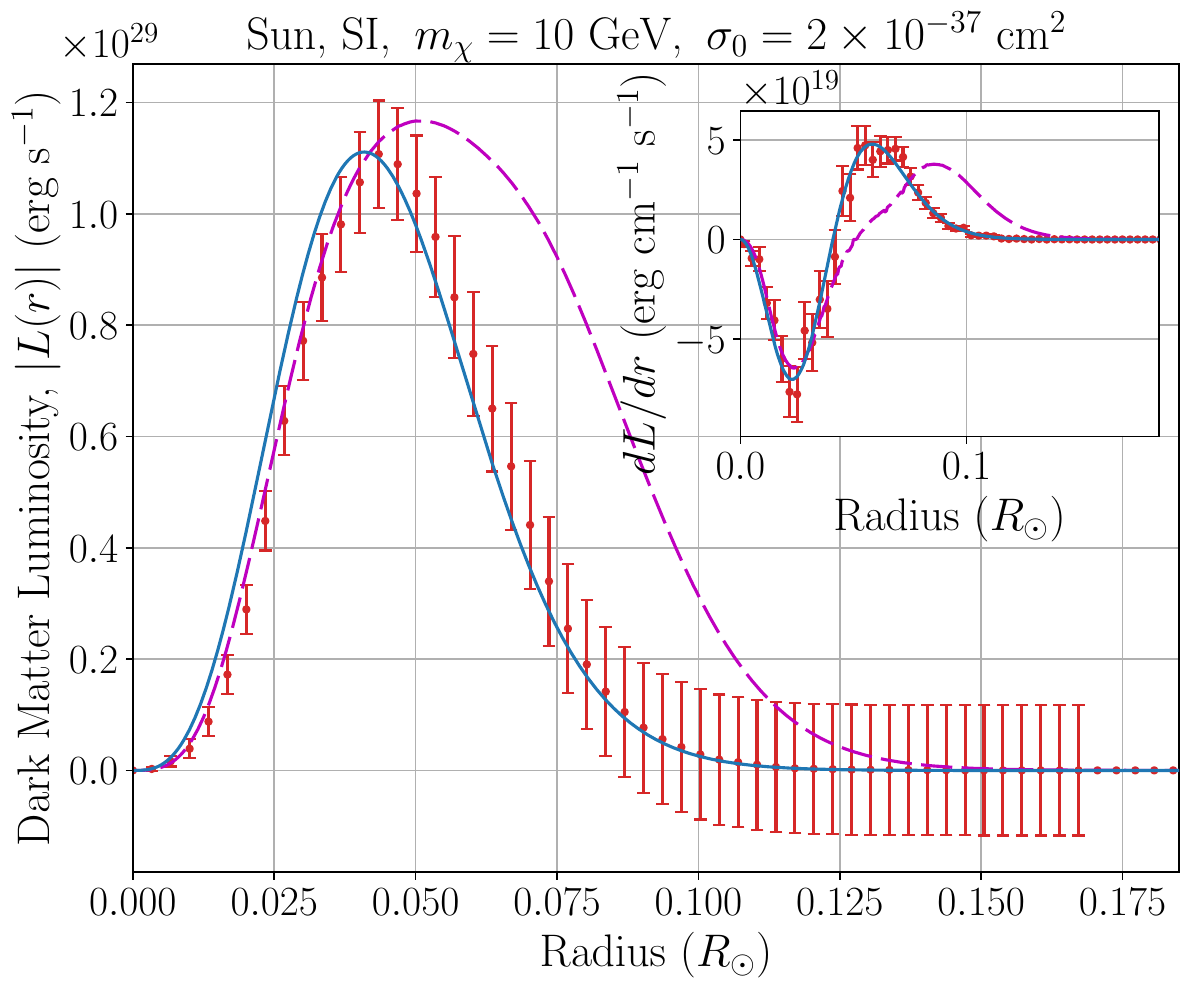}    
    \includegraphics[width=\columnwidth]{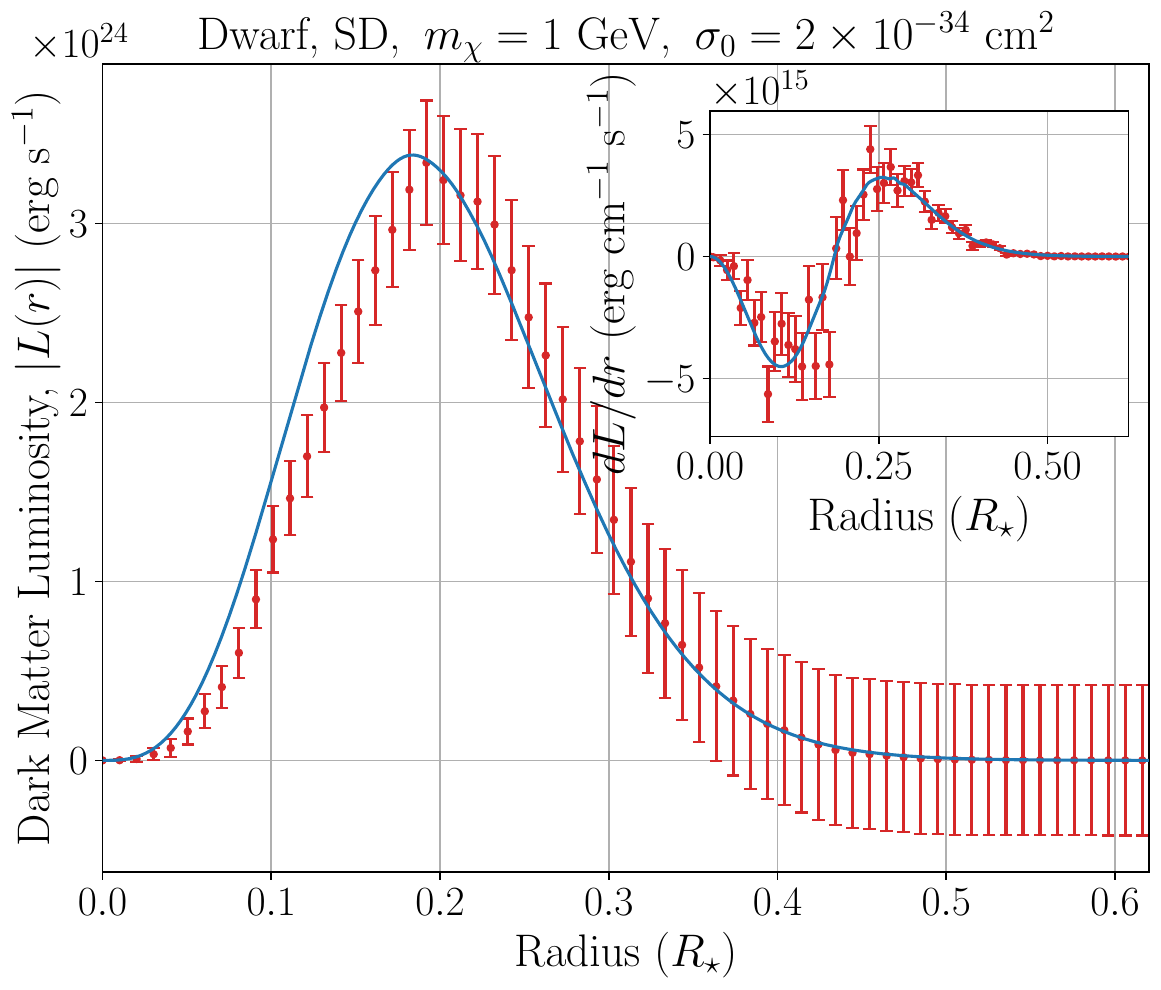}\includegraphics[width=\columnwidth]{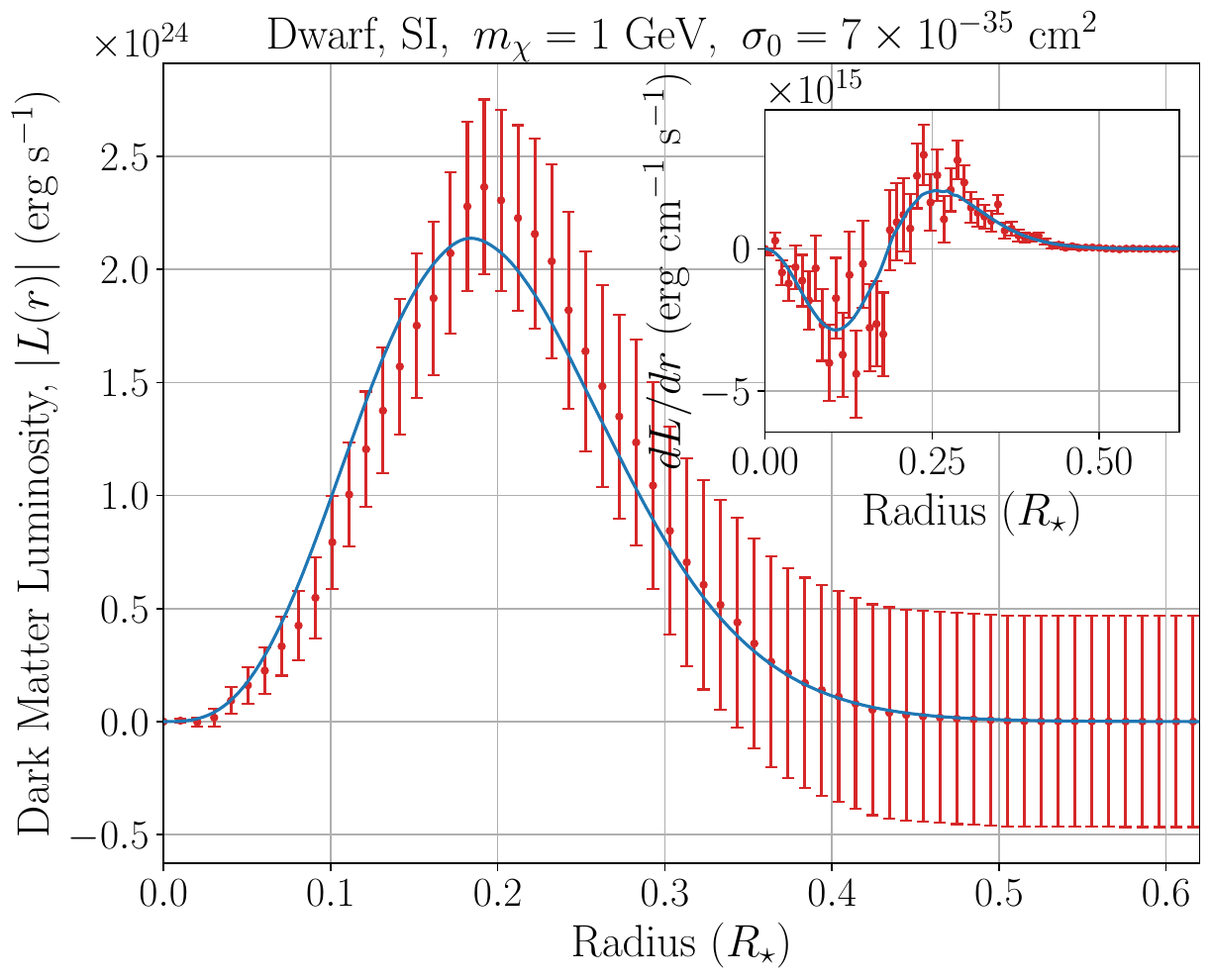}\\
            \includegraphics[width=\columnwidth]{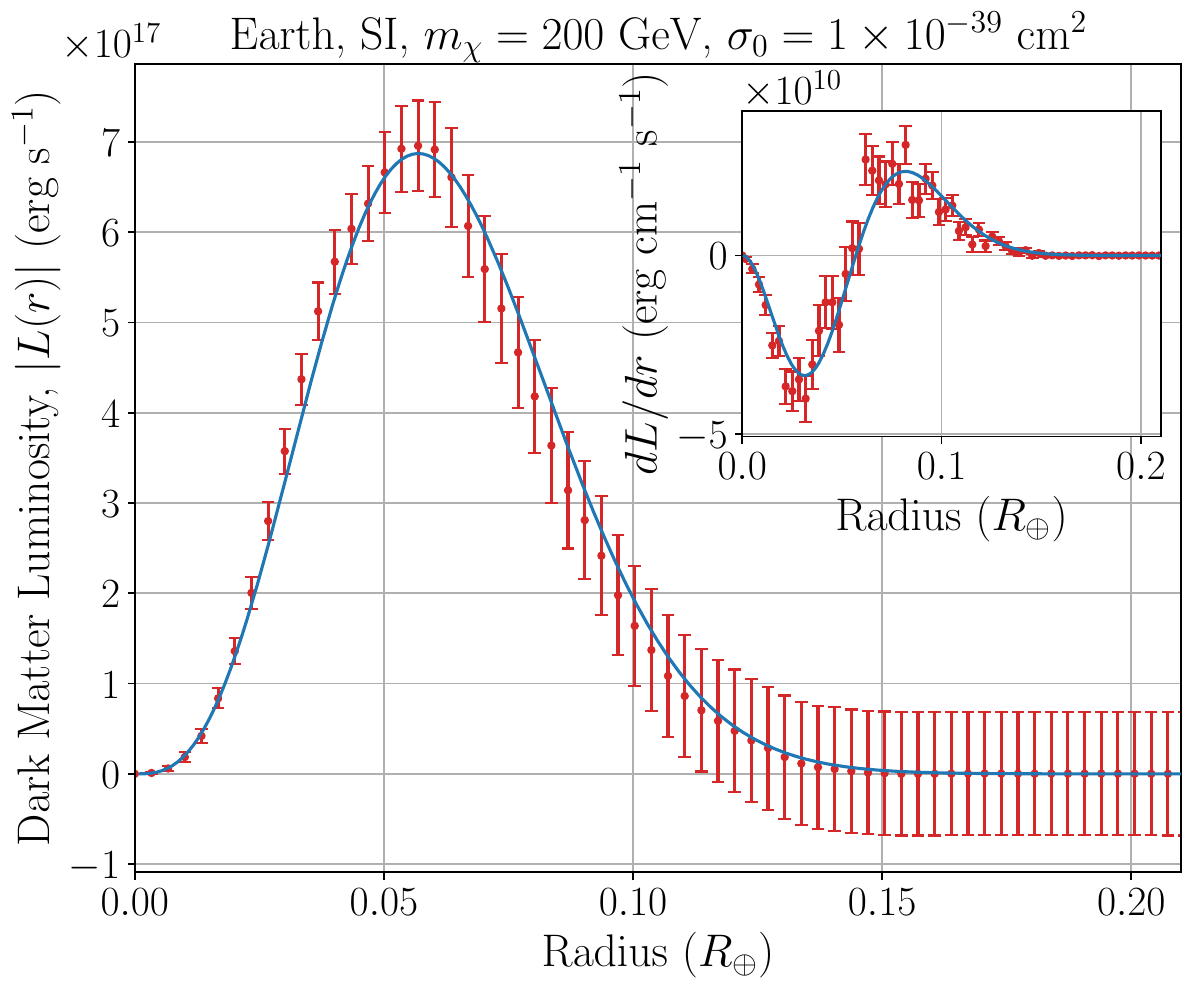}
    \caption{Luminosity $|L(r)|$ and transported energy $dL/dr$ (insets) in the Sun (top), a 0.01 $M_\odot$ brown dwarf (middle) and the Earth (bottom) generated by the \cosmion simulations (data points), and modeled by Eq.~\eqref{eq:SPrescale} (blue solid) with the values of $A$ in Tab.~\ref{tab:params}. We also show the predictions of Gould \& Raffelt \cite{Gould1990} for the Sun. Here, we have taken a DM density per baryon of $n_\chi/n_b=10^{-15}$. }
    \label{fig:SunLuminosity}
\end{figure*}
\begin{figure}[ht!]
    \centering
    \includegraphics[width=\columnwidth]{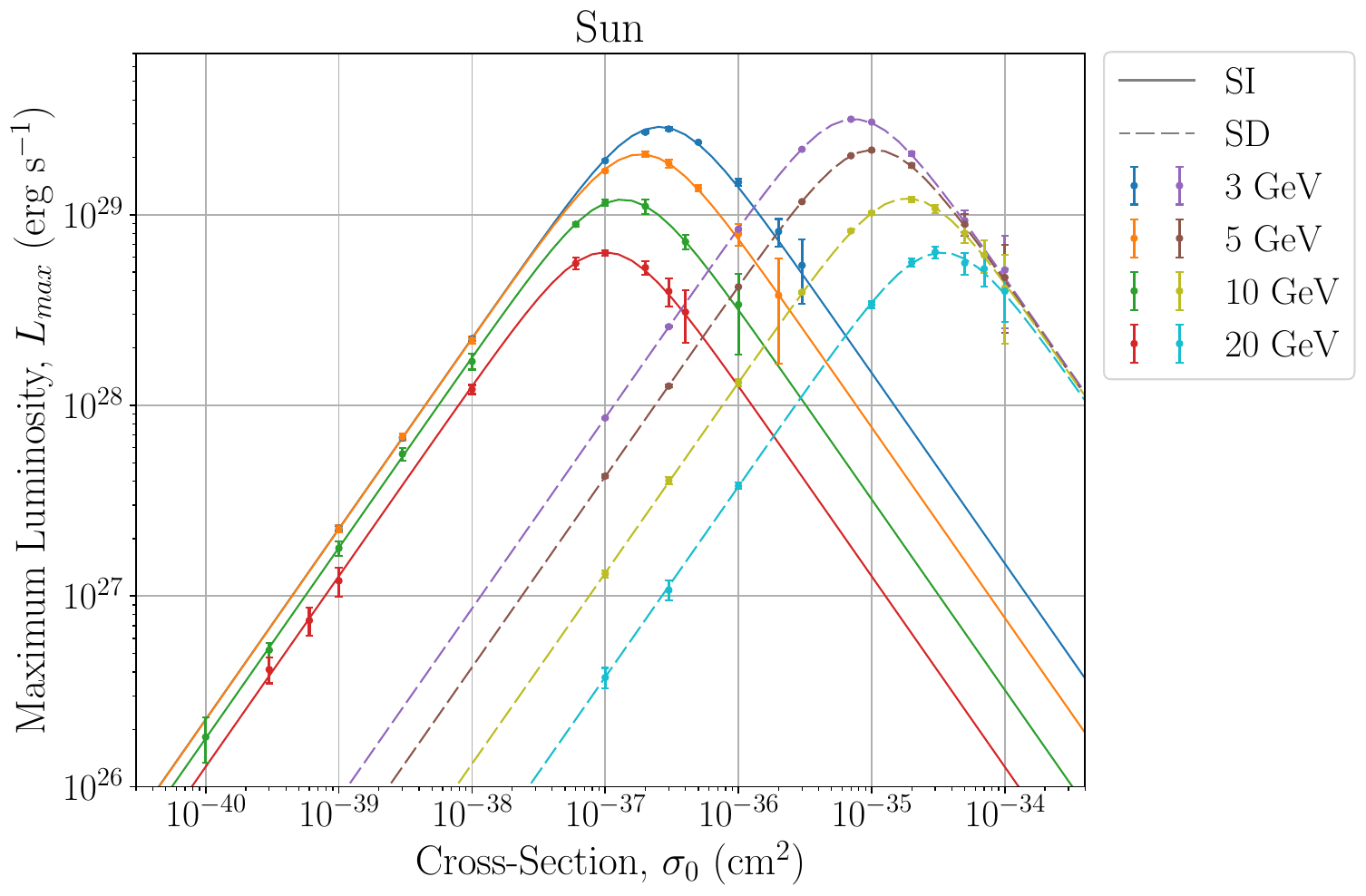}\\
     \includegraphics[width=\linewidth]{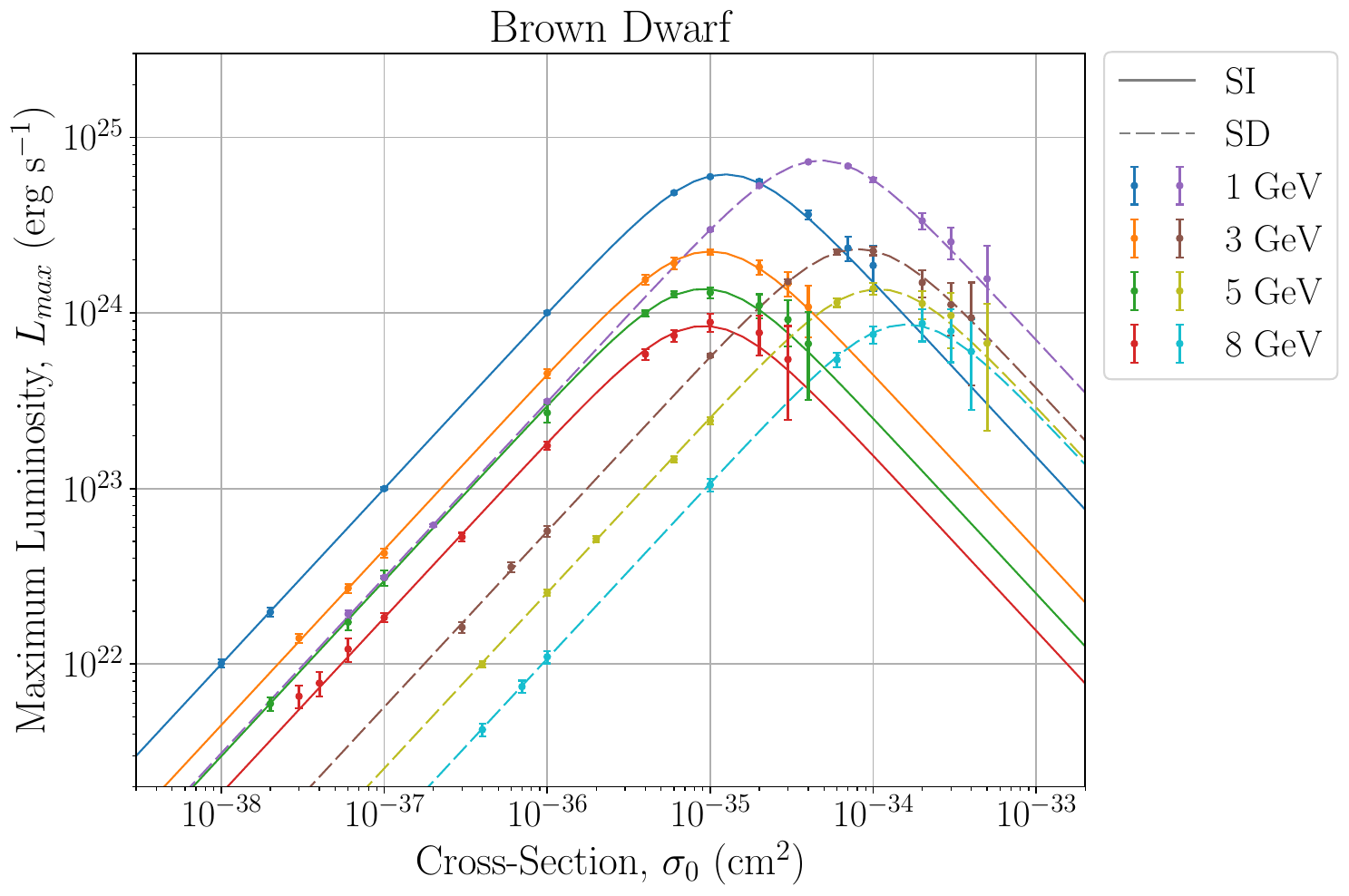}\\
        \includegraphics[width=\columnwidth]{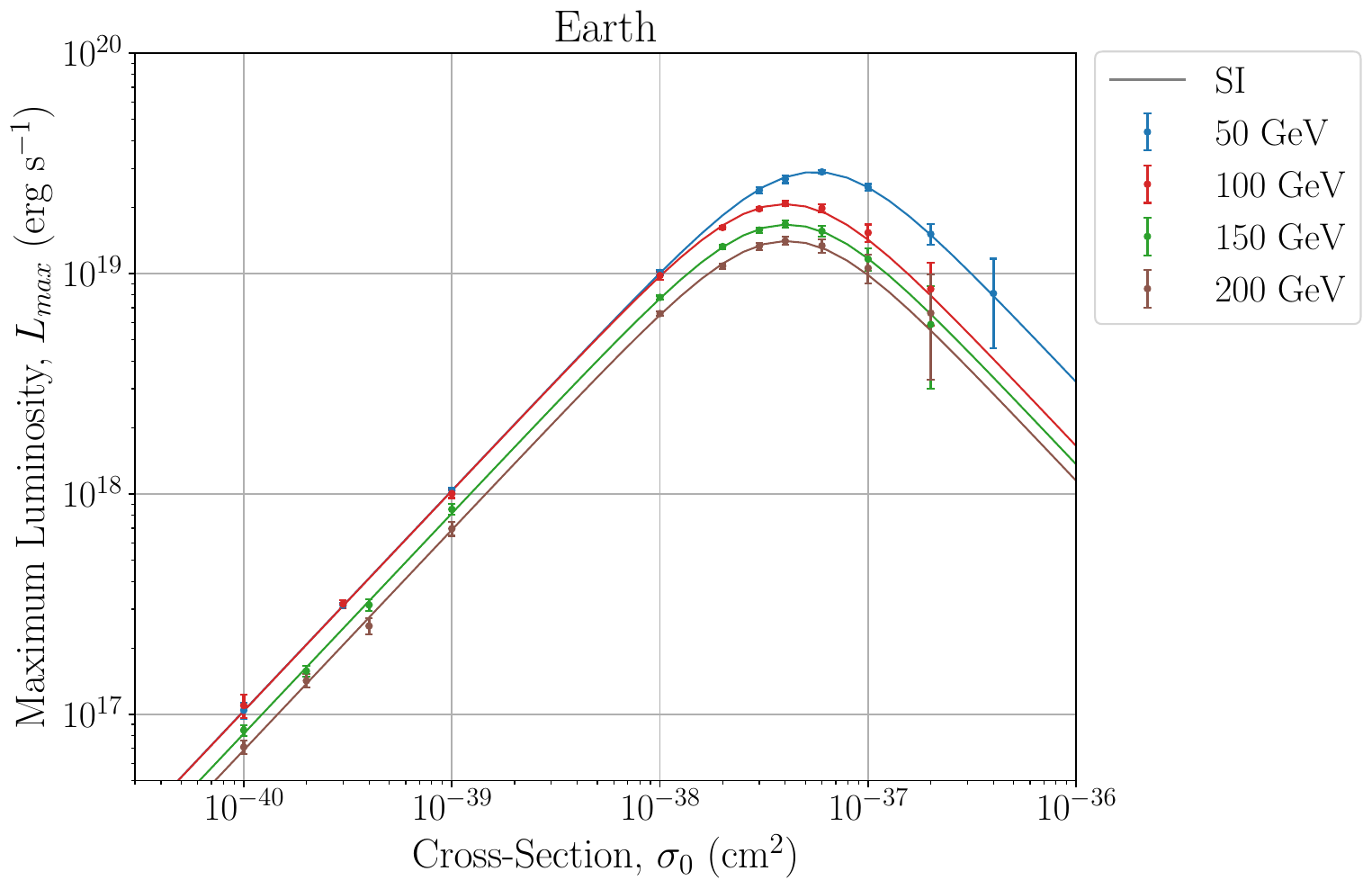}
    \caption{Maximum transported luminosities $L_{max}$ due to DM-nucleon elastic scattering as a function of the cross section for a variety of DM masses, in the Sun (top), a brown dwarf (middle) and the  Earth (bottom). The data points with error bars represent the results of our Monte Carlo simulations in a realistic gravitational potential, for both spin-dependent (SD) and spin-independent (SI) interactions. For the Earth, we only show the SI result due to its composition. The solid curves represent the analytic predictions made by our calibrated Spergel \& Press model \eqref{eq:SPrescale}.}
    \label{fig:SunTransition}
\end{figure}

\section{Heat conduction by weakly-interacting particles}
\label{sec:review}
The phase space distribution $F(\vec{v},\vec{r},t)$ of diffuse  weakly interacting particles in a gravitational potential is formally governed by the Boltzmann collision equation (BCE)
\begin{equation}
\label{eq:BCE}
DF = l^{-1}_{\chi}CF.
\end{equation}
Assuming spherical symmetry, $D = \partial_t + \vec{v} \cdot \nabla_r + \vec g(r)\cdot \nabla_v$ is a differential (Liouville) operator encoding the effects of diffusion and gravity, $l_{\chi}$ is the typical interscattering distance, and $C$ is a collision operator such that $CF(\vec{v},\vec{r},t)$ encodes all collisions to and from velocity $\vec{v}$ at position $\vec r$ and time $t$. Assuming the DM is sufficiently diffuse such that DM-DM scattering events can be neglected, $C$ is linear with a form that depends explicitly on the kinematic structure of the DM-SM interaction. Provided that the collisional time scale is significantly smaller than that over which the structure over the astrophysical body changes, $C$ is time independent and $F$ reaches steady state i.e. $\partial_t F = 0$. 

Once the DM phase space distribution is obtained, the dark matter  heat deposition/removal per unit of stellar material $\epsilon(r)$, or its integral, the effective luminosity
\begin{equation}
    L(r) = \int_0^r r^2 \rho(r) \epsilon(r) dr,
    \label{eq:lumdef}
\end{equation}
may be computed.

Whilst this equation is not in general analytically tractable, 
two approximate solutions have been developed for different limits of $l_\chi$, expressed via the Knudsen number $K = l_\chi(r = 0)/r_\chi$, where $r_\chi$ is the scale radius---the radius at which the distribution of DM confined to a constant density, isothermal sphere with the same  temperature $T_c$ and density $\rho_c$ as the core of the astrophysical body, would peak: 
\begin{equation}
    r_\chi=\sqrt{\frac{3k_BT_c}{2\pi G\rho_cm_\chi}}~.
    \label{eq:scale}
\end{equation}

These two formalisms are:
\begin{enumerate}
\item \textbf{Spergel \& Press (SP):} In the weakly-interacting  ($K \gg 1$) regime (the so-called \textit{Knudsen limit}), the DM is effectively isothermal and heat transport can be modeled as the interaction between two heat baths at different temperatures. The DM temperature $T_\chi$ in this case is the average of the nucleon temperatures with which the DM is in thermal contact with, weighted by the local interaction rate. This solution was mainly developed by  Spergel \& Press (SP) \cite{Spergel1985EffectInterior}, and characteristically becomes weaker with increasing mean free path, as the interaction probability becomes suppressed.  The number density in the SP approach is 
\begin{equation}
    n(r)_{\chi,\mathrm{iso}} = N_\mathrm{iso}e^{-m_{\chi}\phi(r)/T_{\chi}},
    \label{eq:nofr}
\end{equation}
 where $N_\mathrm{iso}$ normalises the distribution, $\phi(r)$ is the gravitational potential, and $T_{\chi}$ is a temperature determined by the condition that $L(R_\star) = 0$. The transported energy is modeled as
 \begin{widetext}
 \begin{equation}
    \epsilon_\mathrm{SP} = \frac{8}{\rho(r)} \sqrt{\frac{2}{\pi }} \frac{m_\chi m_N}{(m_\chi + m_N)^2} n_\chi(r) n_N(r) \sigma_\mathrm{tot}k_B\left(T_\chi-T(r)\right) \left(\frac{k_BT(r)}{m_N}+\frac{k_BT_\chi}{m_\chi}\right)^{\frac{1}{2}},
  \label{eq:SP}
\end{equation}
\end{widetext}
where $\sigma_\mathrm{tot} = \int (d\sigma/d\cos\theta) d\cos\theta = 2\sigma_0$. Equivalent expressions for cross sections that depend on momentum or velocity can be found in Ref.~\cite{Banks:2021sba}.  Ref.~\cite{Gould1990} showed by numerical simulation that the SP approximation actually overpredicts luminosities in the Knudsen regime by approximately a factor of 2, which was confirmed in Ref.~\cite{Banks:2021sba}.

\item \textbf{Gould and Raffelt (GR):} In the \textit{Local Thermal Equilibrium} (LTE, $K \ll 1$) limit, stronger interactions may lead to more scattering, but as a consequence heat transport generally remains ``local'' as DM will typically not travel far before depositing or gaining kinetic energy. The standard treatment of this regime as implemented in modern works, was developed by Gould \& Raffelt  \cite{Gould1990}. The approach relies on a series expansion of Eq.~\eqref{eq:BCE}, followed by a multipole expansion in which the dipole component of $F(v,r)$ is assumed to be responsible for heat conduction. The radial dark matter distribution obtained in this approach is 
\begin{align}
\label{eq:conductiondist}
n_{\chi,\mathrm{LTE}}(r) =& n_{\chi}(0)\left[\frac{T(r)}{T(0)}\right]^{3/2} \nonumber \\ & \times e^{-\int_{0}^{r}dr'\frac{k_\mathrm{B}\alpha(r')\frac{dT(r')}{dr'}+m_{\chi}\frac{d\phi(r')}{dr'}}{k_\mathrm{B}T(r')}}.
\end{align} 
A further ad-hoc modification is then required to account for the breakdown of the dipole approximation at lower radii. Practical implementations typically also include a ``Knudsen correction'' to suppress the luminosity as the $K \ll 1$ assumption breaks down. Since we will not make further use of it in this work, we refer the reader to Ref.~\cite{Banks:2021sba} for a full description of the GR approach. 
\end{enumerate}
Whilst the above frameworks provide an analytic handle on DM-mediated heat transport in specific regimes, the only way to exactly solve to the BCE in an arbitrary density distribution $n_{nuc}(r)$ and temperature gradient $T(r)$ is via a direct Monte Carlo (MC) simulation of the random walk of an interacting DM particle. By ergodicity, the particle traces out the full steady-state phase space distribution of the ensemble over a sufficiently large time $t \gg l_\chi/\langle v \rangle$. In Ref. \cite{Banks:2021sba} we developed such a simulation, building upon the MC methods of Nauenberg \cite{Nauenberg1987} and Gould \& Raffelt \cite{Gould1990,Gould1990CosmionLimit}.  We showed that, contrary to the common assumption in the literature, the SP solution provides a more robust description of heat transport by weakly interacting particles, provided a small correction that depends weakly on the kinematic structure of the DM-SM interaction. The simulations in Ref. \cite{Banks:2021sba} were limited to the study of SD interactions in two specific scenarios: an `idealized' toy stellar model matching Refs.~\cite{Gould1990,Gould1990CosmionLimit} ($m_\chi = 1$ kg, $R_\star = 2.4$ m), and a realistic solar model using temperature, density and composition data from the AGS05 Standard Solar Model. The latter formed the first numerical study of DM mediated thermal conduction in a realistic astrophysical environment. In both cases however, the gravitational potential was modeled \`a la Gould \& Raffelt to be that of a simple harmonic oscillator (SHO). Modeling the potential in this way allows the trajectories to be calculated analytically, offering significant advantage in terms of computational accuracy and speed. Whilst a good approximation for particles that are confined to the approximately constant density of the solar core,  it breaks down at higher radii and in other astrophysical bodies. In addition, although our simulations covered a variety of different cross section scalings with velocity and momentum, they were nonetheless restricted to spin-dependent scattering on hydrogen only. The key result of Ref.~\cite{Banks:2021sba} was that a modified version of the SP treatment was sufficient to reproduce the heat transport profile of dark matter in a stellar object, with an SHO potential and assuming scattering only with hydrogen. That is, the luminosity due to dark matter is well-described by:
\begin{equation}
 L(r) =  \frac{A}{1+(K_0/K)^2}L_\mathrm{SP} (r),
    \label{eq:SPrescale}
\end{equation}
where $L_{\mathrm{SP}}$ can be found via Eqs. \eqref{eq:lumdef} and \eqref{eq:SP}, $K_0 \approx 0.4$ represents the Knudsen transition from LTE to isothermal behavior, and $A = 1/2$. 

In this work, we extend our simulation to numerically integrate trajectories in the real gravitational potential of the Sun and other astrophysical bodies. We also examine spin-independent interactions, i.e. heat transport when scattering with many species. 
\section{Monte-Carlo Methodology}
\label{sec:mc}
We outline the Monte Carlo algorithm implemented in the \cosmion code, focusing specifically on the modifications with respect to the procedure presented in Ref.~\cite{Banks:2021sba}.

The software is written in modern Fortran for speed and backward compatibility, and consists of a main program \cosmion, a \texttt{star} module that pre-loads and computes stellar properties, and a set of \texttt{walk} functions and subroutines that encode the initialization, propagation and collisions of the DM particle. 

The random walk algorithm is constructed from the following steps:
\begin{enumerate}
    \item \textbf{Select an initial position and velocity}. Positions are  drawn randomly from the radial distribution in Eq. \eqref{eq:nofr}, and velocities are drawn from a three-dimensional Maxwell-Boltzmann distribution at temperature $T_\chi$.
    \item \textbf{Determine the next point of collision.} This is achieved by randomly choosing an optical depth $\tau$, where
    \begin{equation}
\label{eq:dtau}
d\tau = \sum_{i} \omega_{i}(\vec{v}) dt.
\end{equation}
The collision probability per unit time is 
\begin{equation}
\label{eq:rate}
\omega_{i} = \int d^3 \vec{u} \sigma_{i}(\left|\vec{v}-\vec{u}\right|)n_{i}(r)|\vec{v} - \vec{u}|f(\vec{u};r),
\end{equation}
where the sum is over isotopes $i$. The resulting value of $t(\tau)$ determines the location of the next collision. For a constant cross section, $\omega_i = 2\sigma_{0}n_iv_{T}\sqrt{\mu}\left[\big(y+ \frac{1}{2y})\mathrm{erf}{(y)}+\frac{1}{\sqrt{\pi}}\exp{(-y^2)}\right]$, where $v_T^2 = 2k_\mathrm{B}T/m_{\chi}$ and $y^2 = |\vec{v}/v_T|^2/\mu$. Expressions for this integral for non-constant DM-nucleus interaction cross sections are provided in Table I of Ref. \cite{Banks:2021sba}.
\item \textbf{Integrate the particle's trajectory} to the collision point. In our previous work \cite{Banks:2021sba}, as per Refs. \cite{Gould1990,Gould1990CosmionLimit}, we assumed a simple harmonic potential $\phi(r) = \Omega^2r^2/2$, with $\Omega = \sqrt{4\pi G_N \rho_c/3}$, such that the trajectories could be determined analytically.  Here, we instead use the full, real potential of an arbitrary star $\phi(r)$ and thus explicitly need to integrate in order to find the next collision point.   We use spherical symmetry to eliminate the azimuthal angle, and conservation of angular momentum $\vec L$ to solve for the remaining angular velocity.  Defining
\begin{align}
    \hat r &= \vec r/|r|, \\
    v_r &=  | \vec v \cdot  \hat r|, \label{eq:vrdef}\\
    v_\theta &= |\vec v - v_r \hat r| , \label{eq:vtdef}
\end{align}

with $\vec \ell = \vec L/m = \vec r \times \vec v$, we obtain the following  equations of motion:
\begin{align}
    \dot r & =  v_r \label{eq:rdot},\\
    \dot v_r &= -g(r) + \frac{\ell^2}{r^3}, \label{eq:vdot}
\end{align}
where $g(r) = -\partial \phi/\partial r$ and $\ell \equiv |\vec \ell|$ is conserved between scattering events.
From the conservation of angular momentum, $v_\theta = \frac{\ell}{r}$.

The coordinate system can always be rotated so that propagation is in a plane, and the azimuthal coordinate $\phi$ can be ignored. Thus, the total velocity $v$ can be obtained via
\begin{equation}
    v^2 = v_r^2 + v_\theta^2~. 
\end{equation}
As we are integrating the trajectory up to a specific optical depth $\tau$, we use $\tau$, rather than time $t$, as the dependent variable. Combining Eqs. \ref{eq:rate}, \ref{eq:rdot} and \ref{eq:vdot}, we obtain the following system of equations:
\begin{align}
    \frac{dt}{d\tau} &= \omega^{-1}(v),  \\
    \frac{dr}{d\tau} &= \frac{dt}{d\tau}  \dot r,   \\
        \frac{dv_r}{d\tau} &= \frac{dt}{d\tau}  \dot v_r.  \label{eq:rk3}
\end{align}
This is a set of coupled ODEs which we solve using a Runge-Kutta-Fehlberg (RKF45) integrator.

If at any point the integrator determines that the particle will reach the boundary of the star before scattering again, the propagation is aborted. If the particle's speed is above the local escape velocity, the event is recorded as an evaporation event, and a new particle is spawned. Otherwise, the particle's exit and re-entry is computed in two steps. First, the optical depth to the edge of the star from the last collision is found, and the particle is propagated towards the boundary in order to record the time taken during this stage. Next, the particle's Keplerian orbit outside the star is computed, to determine the time elapsed prior to re-entry. The particle is then reinitialized on the boundary at the point of re-entry, and the simulation proceeds as normal. Further details on the propagation outside the star are given in Appendix \ref{app:kepler}.
\item  (For spin-independent scattering) \textbf{Select a nuclear species with which to scatter}. Whereas our previous work in Ref.~\cite{Banks:2021sba}  only considered spin-dependent collisions with hydrogen, the present simulation allows for the investigation of both spin-dependent and spin-independent interactions.  Owing to the much lower abundance of other nuclear species with nonzero spin,  we once again restrict the spin-dependent case  to scattering with hydrogen only. Whilst other nuclei can be important for SD scattering in some situations, e.g. in rocky planets such as the Earth; these are not currently accounted for in the current release of \texttt{cosmion}. For spin-independent interactions, we allow for scattering with the most abundant nuclear species in each celestial body. Modifications to the particle propagation for spin-independent interactions are already taken into account by the collision probabilities $\omega_i$ used to compute the optical depth in Eq.~(\ref{eq:dtau}). Spin-independent simulations take much longer to run in general, as each step of the integration of the optical depth $\tau$ during particle propagation requires the computation of the interaction rate $\omega_i$ for each species. To reduce the runtime of these simulations, we have developed a selection mechanism to consider only a subset of the total nuclear species with which the dark matter particle can collide. The user may specify the required precision, which will select the species based on their interaction cross-section and abundance. More details on this procedure can be found in Appendix \ref{app:collisionrate}.

\item \textbf{Draw a nuclear velocity}, sampling a speed $u$ and scattering angle $\vartheta$ from the distribution
\begin{equation}
\left(v^2 + u^2 - 2uv \cos \vartheta\right)^{1/2} e^{-m_i u^2/2k_BT}.
\end{equation}
To do this we implement the sampling algorithm outlined in Ref.~\cite{romano2018improved}.\footnote{Noting a missing square root in the conditional expression of Algorithm 1: $\xi_6 < \sqrt{y^2 + z - 2xy\mu}/{(y + x)}$.}

\item \textbf{Perform the collision} in the centre-of-momentum frame, and randomly assign new velocities to the dark matter particle according to the differential cross section.
\end{enumerate}
Steps 2-6 are then repeated a large number of times ($N \sim 10^6- 10^9$, depending on the interscattering distance) to obtain a converged phase space distribution, as well as a record of energy transfer as a function of radius.

At each collision point the position, and the ingoing and outgoing DM velocities are recorded. From these, the energy transferred per collision $i$ is computed:
\begin{equation}
    \Delta E_i = \frac{1}{2}m_\chi (v_{out}^2 - v_{in}^2)~.
\end{equation}
This can then be interpreted as in Ref. \cite{Banks:2021sba}.

If we want the equilibrium distribution versus radius, recording the samples at each collision point provides a biased sample. To correct this, we note that the recorded collision points have been sampled with a probability
    \begin{equation}
        \Gamma_i \propto n_\chi(r) \omega(\vec v,\vec r)
    \end{equation}
    such that each point can be weighted by $\omega^{-1}$ to approximate the true distribution $n_\chi(r)$. This weight is provided as an output of the \texttt{cosmion} code.

\begin{figure*}[ht!]
    \centering
    \includegraphics[width=\columnwidth]{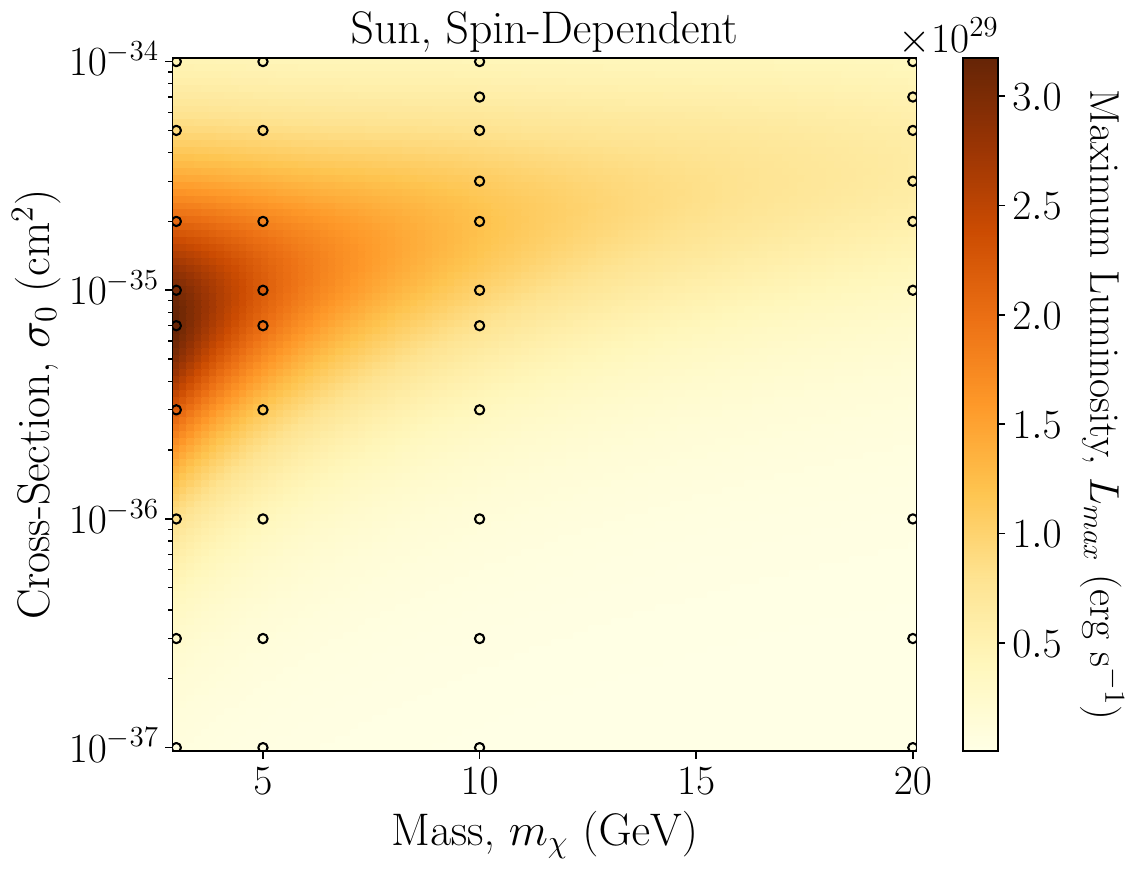}\includegraphics[width=\columnwidth]{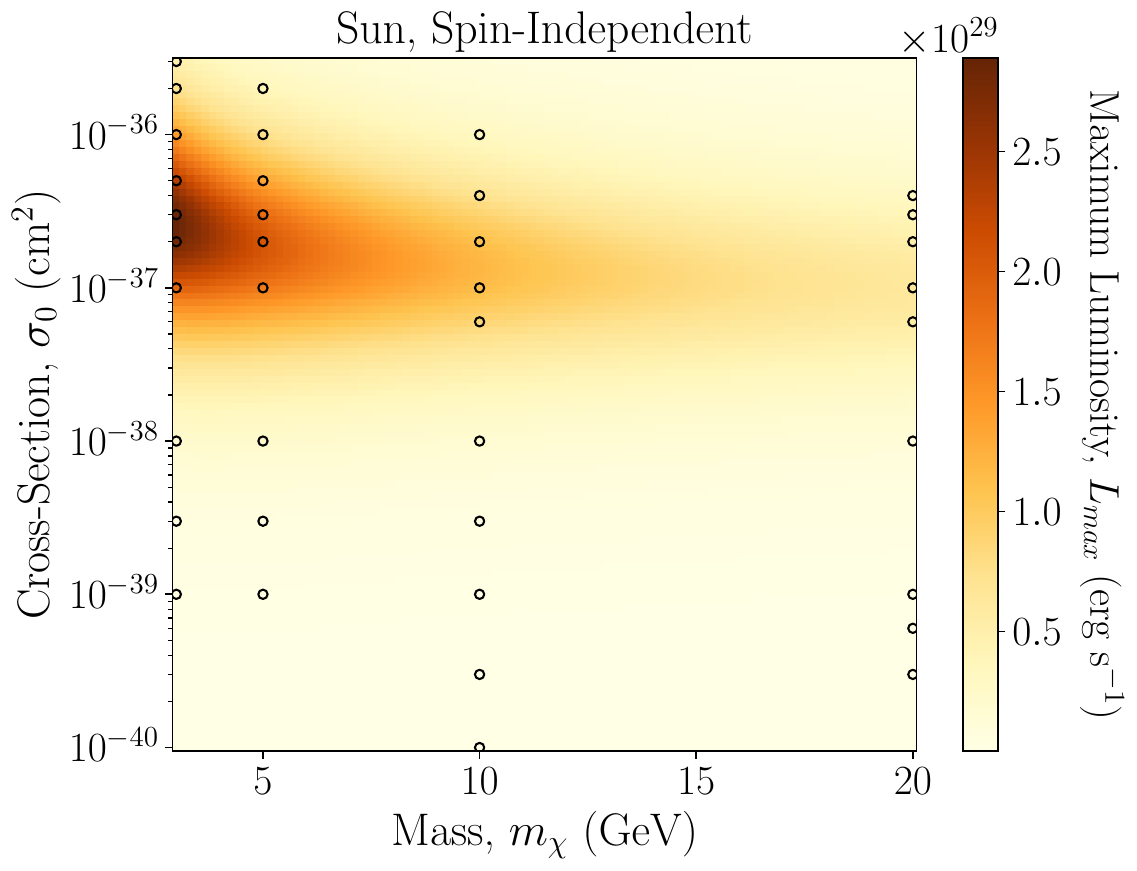}
        \includegraphics[width=\columnwidth]{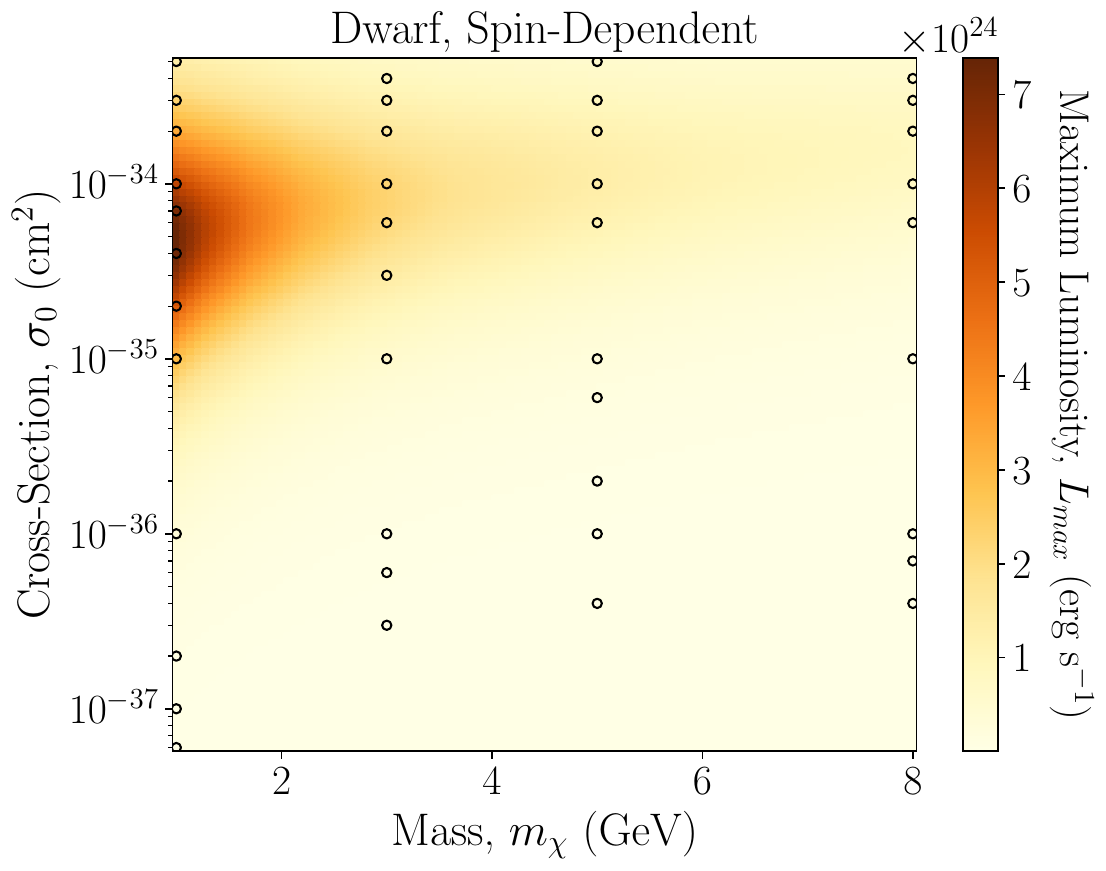}\includegraphics[width=\columnwidth]{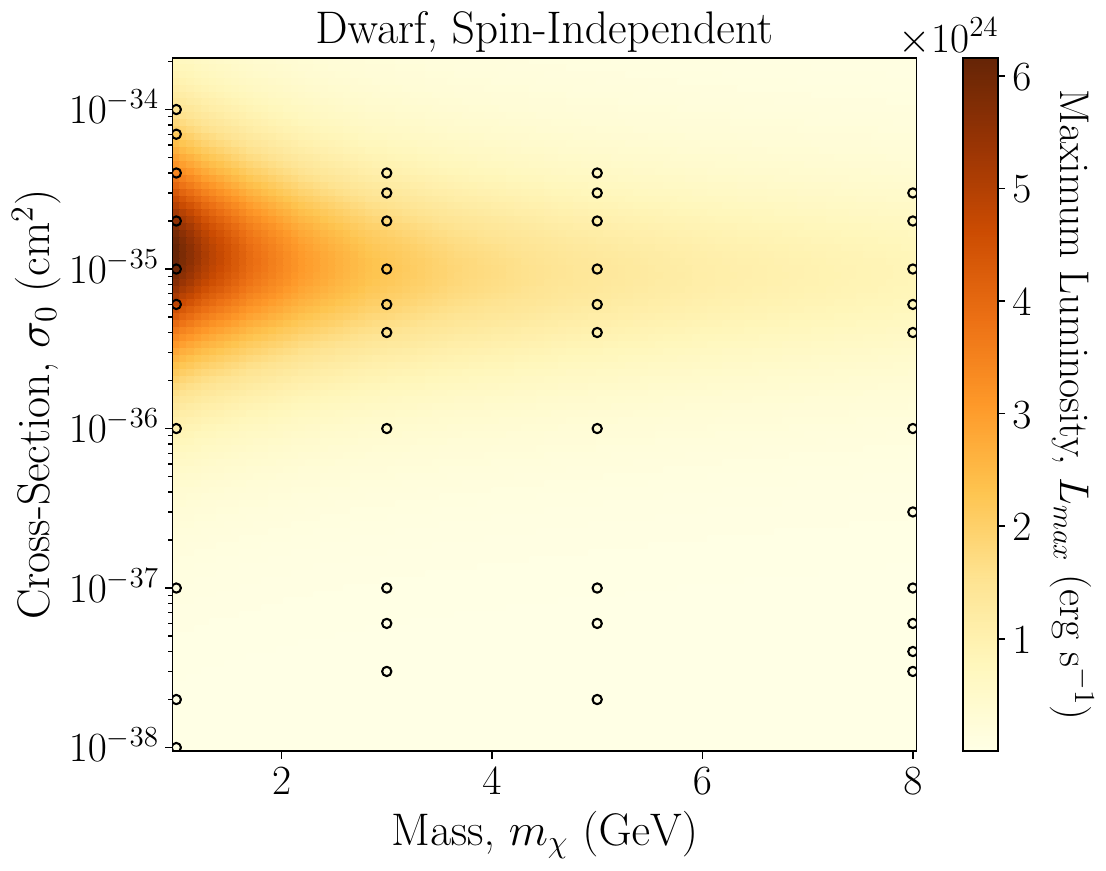} \\
            \includegraphics[width=\columnwidth]{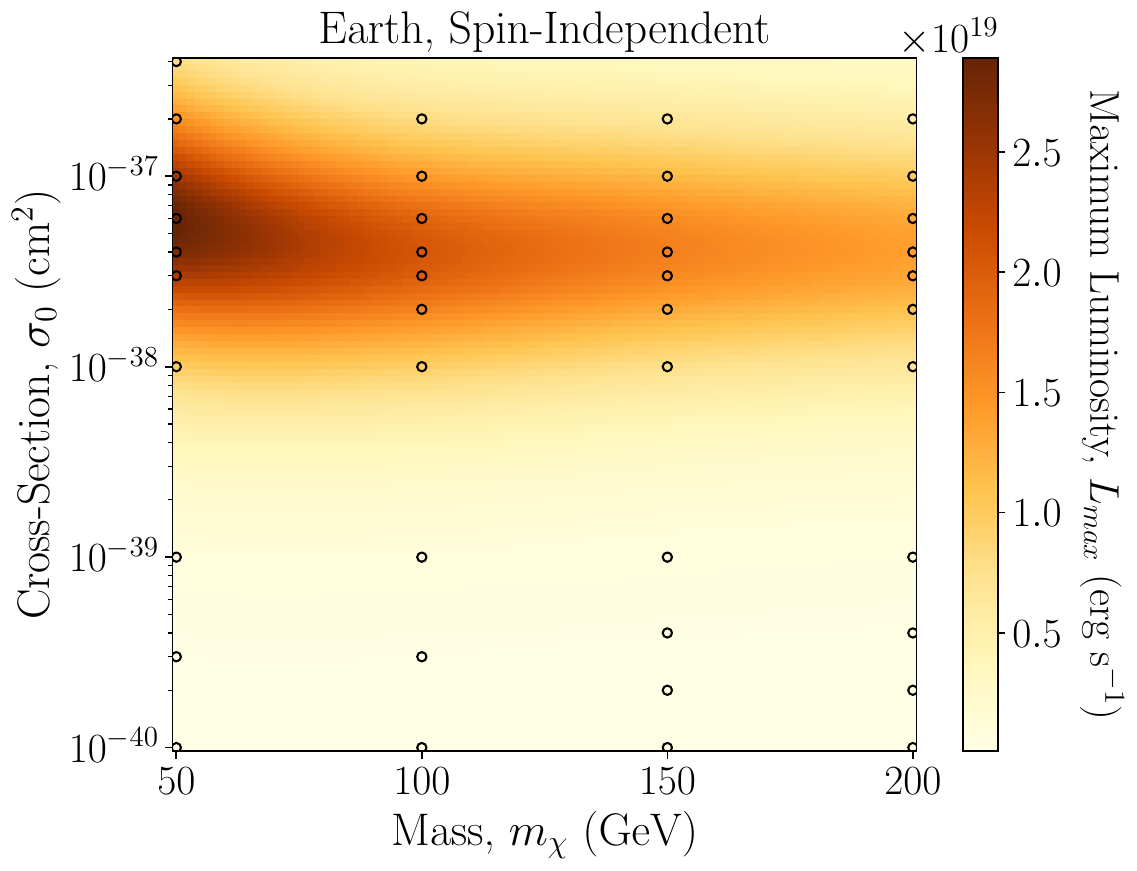}
    \caption{Peak luminosities of a dark matter distribution in the Sun (top), brown dwarf (middle) and Earth (bottom), as a function of particle mass $m_\chi$ and constant interaction cross-section $\sigma_0$. The results were computed via the calibrated Spergel \& Press formalism \eqref{eq:SPrescale} in a realistic gravitational potential, with a relative number abundance of dark matter particles to solar baryons of $n_\chi/n_b=10^{-15}$. The formalism's parameters $\alpha$ and $K_0$ were determined by way of a monotonic cubic interpolation between the values in Table \ref{tab:params} across the particle mass range. The black rings represent the simulation data points that were used to calibrate the parameters for the formalism.}
    \label{fig:SunHeatmapSD}
\end{figure*}
\section{Simulation results}
\label{sec:results}
In this section, we present results of our simulations and contrast them with analytic approximation methods. We will show the effects of including a realistic gravitational potential, spin-independent scattering with multiple nuclear species, and evaporation. We will show the effects of heat transport in a realistic Sun, brown dwarf, and Earth model.

The Solar model used in this work is the B16 AGSS09met standard solar model from Ref.~\cite{Vinyoles2017}\footnote{For the solar data file used in this work, visit \href{https://www.ice.csic.es/personal/aldos/Solar_Models.html}{https://www.ice.csic.es/personal/aldos/Solar\_Models.html}.} The 0.01 $M_\odot$ brown dwarf model was generated using MESA~\cite{Paxton2011, Paxton2013, Paxton2015, Paxton2018, Paxton2019, Jermyn2023}. The temperature and density profiles of the Earth were derived from Refs. \cite{Dziewonski1981,Jaupart2015,Arevalo2009,Earle2015} as compiled in Ref. \cite{Acevedo2021}, while the compositions of the core, mantle, and crust were obtained from Refs. \cite{Alfe2007,Workman2005,Asimov1954}, respectively. A variety of DM particle masses were simulated in each system, with cross sections ranging over multiple orders of magnitude. Both spin-dependent couplings to hydrogen and spin-independent interactions were considered in the Sun and the brown dwarf, while only spin-independent interactions were used in the Earth due to its relative lack of hydrogen. Each simulation of a particle with a particular mass and cross section in a given system was carried out over $10^6$ -- $10^9$ collisions, with lower cross sections generally requiring fewer collisions, and higher cross-sections requiring more in order to sample the phase space distribution of the particle. The CPU time per collision depends on the cross section: in the LTE regime, these can take below $10^{-5}$ s per collision, whereas longer trajectories in the isothermal regime take longer (0.1 s at $\sigma_{SD} = 10^{-38}$ cm$^{2}$).  Despite this, simulations in the LTE regime end up taking far more CPU time, as exponentially more interactions must be simulated to properly cover the phase space (in other words, the particle takes a long time to move around the star). The software is therefore at its best near the Knudsen transition, where enough statistics can be accumulated without much loss of accuracy. The iteration over nuclear species, in the case of spin-independent scattering, is parallelised with OpenMP; however, the easiest way to speed up evaluation is by launching multiple instances on separate CPUs and combining the outputs.

\subsection{Heat transport}
DM-nucleon scattering with hydrogen only, in a SHO gravitational potential, was explored in detail in Ref.~\cite{Banks:2021sba}. There, it was found that Eq.~\eqref{eq:SPrescale} accurately predicted the heat transport rate across a large range of dark matter masses, cross sections and interaction types. The rescaling factor $A = 1/2$ was in agreement with the conclusions of Ref.~\cite{Gould1990EvaporationSections} who found the SP overestimated transport rates by about a factor of two. We will leave this factor as a free parameter that we fit based on simulation results. Though $A = 1/2$ generally approximates the results quite well, we will find that a slightly lower value generally provides a better fit. Results of this fit are presented in Table \ref{tab:params}, where we also fit for the location of the Knudsen transition, $K_0$, which is robustly found to be between $0.4$ and $0.5$.

Examples of the equilibrium radial distribution are shown in Fig.~\ref{fig:SunRadial} for a dark matter mass $m_\chi = 3$ GeV for spin-dependent (left) and spin-independent (right) interactions. Simulation results are shown in shaded blue, and the different lines represent the SP radial distribution function \eqref{eq:nofr} (Isothermal, red solid line), the GR function \eqref{eq:conductiondist} (LTE, dot-dashed), and the interpolation between the two commonly used in the literature (dashed, green). We note that an additional component as predicted in Ref.~\cite{Leane:2022hkk} would not appear here, since the flux due to capture is not present in the simulation.

Starting with a realistic solar model, we perform heat transport simulations for a range of DM masses, and across $\sim 5$ orders of magnitude in cross section, in order to properly cover the Knudsen transition. Some examples of the transported luminosity $L(r)$ as a function of radius can be found in the top panels of Fig. \ref{fig:SunLuminosity}, for SD scattering (left), and SI scattering (right). Throughout, we assume that the ratio of captured DM to baryons is $n_\chi/n_b = 10^{-15}$. The inset in each of these plots shows the luminosity derivative $dL/dr \propto \epsilon$. In both cases, we show the luminosity near the Knudsen transition. The lines represent predictions from the GR formalism (purple, dashed), and from the calibrated SP formalism (blue). As in Ref. \cite{Banks:2021sba}, we find that the GR formalism does not describe the shape of the heat transport curve very well near the Knudsen transition. We perform a fit to the values of $K_0$ and $A$ in Eq. \eqref{eq:SPrescale} that best fit the maximum luminosity: we first find the value of $A$ that matches simulations in the isothermal regime, and then find the turnaround value of $K_0$ to cover the Knudsen transition. These values exhibit some weak dependence on the DM mass, and differ for spin-dependent versus spin-independent scattering. Tab. \ref{tab:params} provides the best-fit values of these coefficients for masses from 3 to 20 GeV, where heat transport is the most efficient, and therefore the most likely to affect stellar structure. Below these masses we expect evaporation to deplete the DM population, whereas at higher masses, the inefficient heat transfer means that we were not able to obtain converged numerical values. The values of $A$ that we find are consistently around $0.43$, i.e. slightly lower than the factor of $1/2$ mentioned in Ref. \cite{Gould1990CosmionLimit}. As in Ref. \cite{Banks:2021sba}, we attribute this suppression to the fact that the DM temperature distribution is not actually isothermal, but lies somewhat closer to the star's temperature across all radii. 

The top panel Fig. \ref{fig:SunTransition} shows the maximum luminosity obtained in each simulation (data points) across the full range of cross sections, for four different masses. We overlay the calibrated SP prediction for spin-independent (solid) and spin-dependent scattering, showing that the two-parameter model  well describes heat transport across the parameter space.

The middle panels of Fig. \ref{fig:SunLuminosity} show the corresponding heat transport curve in the case of a 0.01 $M_\odot$ brown dwarf. Results here are very similar to those obtained in the Sun, though we observe less variation in the fitted value of $K_0$. The middle panel of Fig. \ref{fig:SunTransition} shows the range of maximum luminosities across the Knudsen transitions for masses from 1 to 8 GeV, again showing good agreement between the parametrization and simulation results.

Finally, the bottom panels of Fig. \ref{fig:SunLuminosity} and Fig. \ref{fig:SunTransition} show results for scattering in the Earth. Here, we only show spin-independent results due to the Earth's composition. Even though the Earth's composition and structure are very different from those of stars, the SP formalism continues to provide an excellent prediction of the transported heat. 

Fig. \ref{fig:SunHeatmapSD} summarises these results, showing the peak luminosity in the $\sigma-m$ plane. The colour map here is made using Eq. \eqref{eq:SPrescale}, with values of $A$ and $K_0$ from Tab. \ref{tab:params}. The circles in each plot show the grid of simulations performed. As expected, heat transport peaks in efficiency at lower masses, when scattering is best kinematically matched with the most abundant target species.

\begin{figure}
    \centering
    \includegraphics[width=\linewidth]{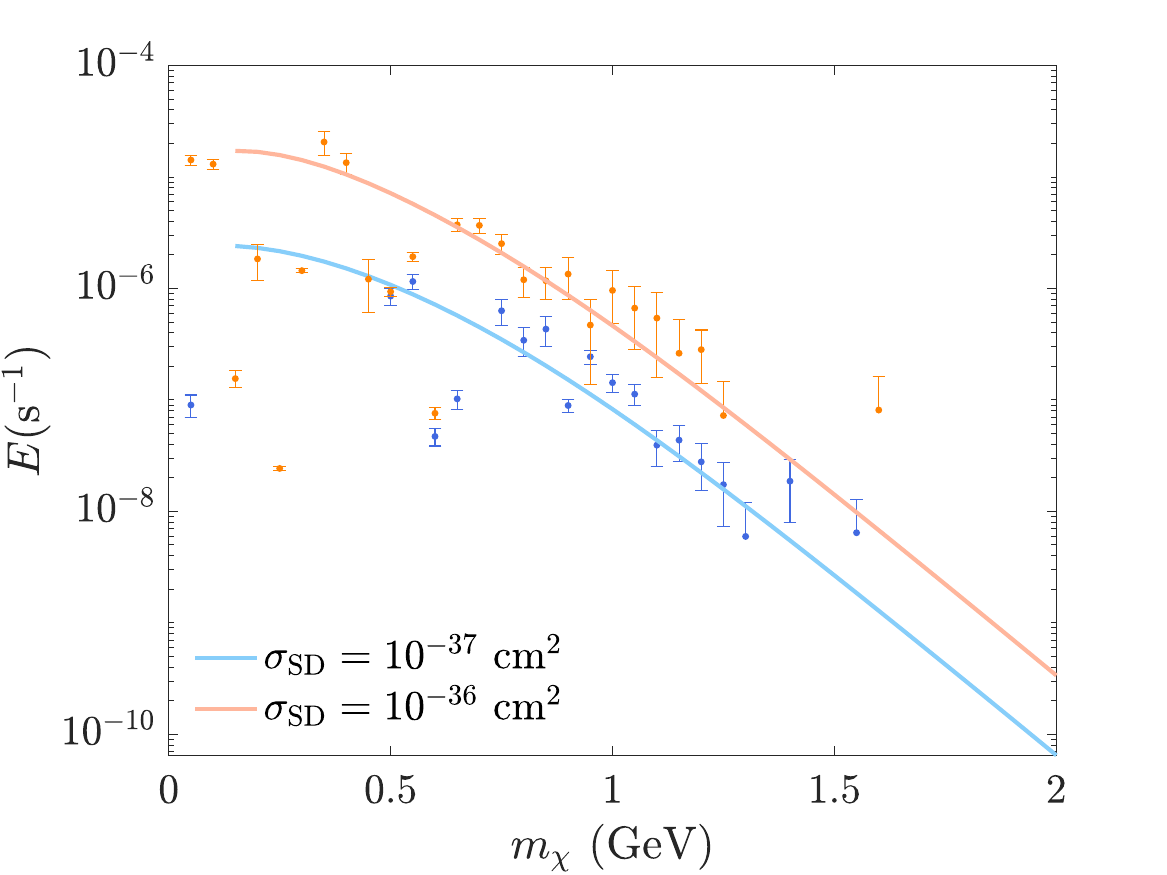} \\
    \includegraphics[width=\linewidth]{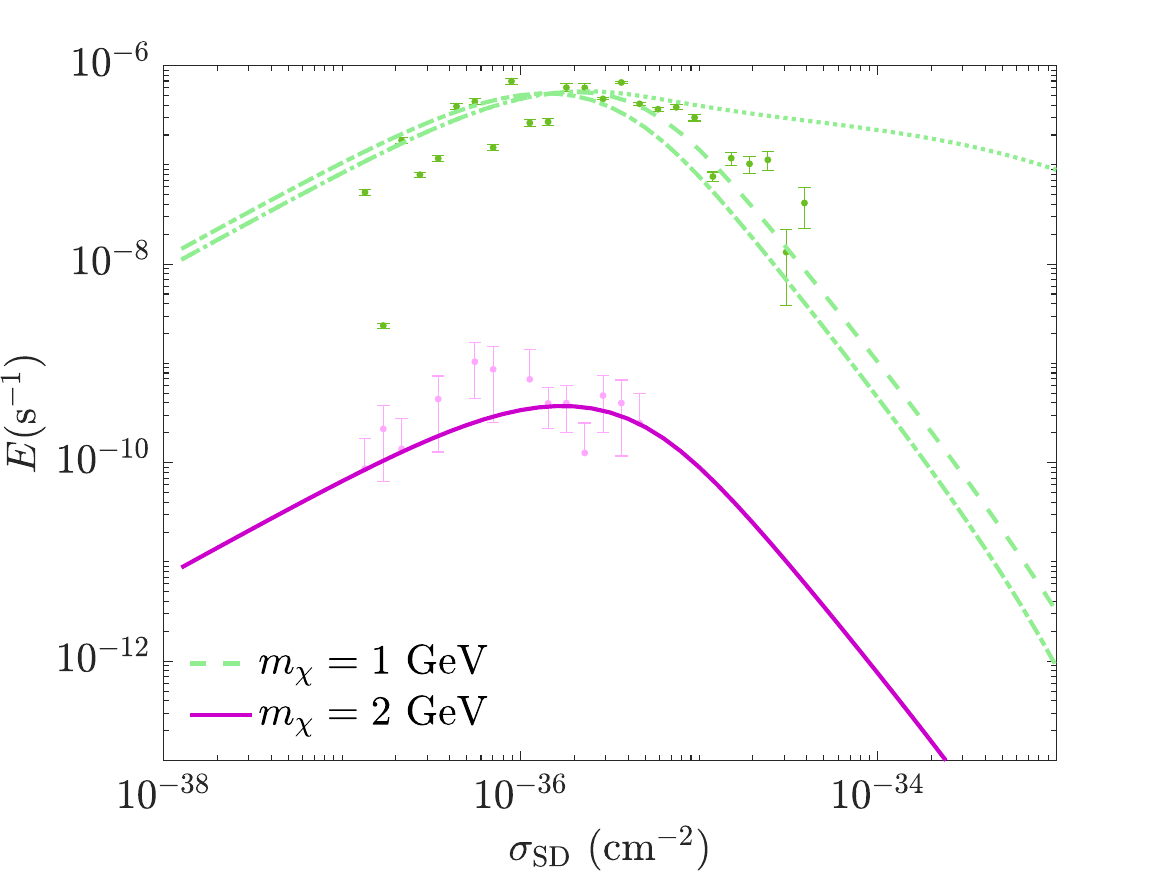} \\
    \caption{Evaporation rate of dark matter from the Sun, as a function of the DM mass (top) and cross section (bottom). Lines show the rates predicted using the approach of Ref. \cite{Busoni2017EvaporationSun}. In the 1 GeV case, we show the predictions using the isothermal approach (dotted), LTE (dot-dashed), and the interpolation suggested in Ref. \cite{Busoni2017EvaporationSun}. For all other cases, we only show the interpolation.}
    \label{fig:evapvsm}
\end{figure}
\subsection{Evaporation}
The evaporation rate can be computed after a set of simulations by dividing the number of evaporation events $N_{\rm evap}$ (followed by respawning of a new particle) by the total residence time recorded by the simulation $t_{\rm sim}$. This is a time-consuming and error-prone process, however. As pointed out e.g. in Ref. \cite{Busoni2017EvaporationSun}, evaporation becomes important when 
\begin{equation}
    \frac{1-e^{-\epsilon t_\star}}{\epsilon t_\star} \sim \frac{1}{2}~,
\end{equation}
where $\epsilon$ is the evaporation rate and $t_\star$ is the stellar age. For $t_\star \sim$ Gyr, this gives relevant evaporation rates of $10^{-10}$ yr$^{-1}$. Running simulations long enough to provide enough evaporation events near the expected evaporation mass (e.g. 4 GeV in the Sun) is thus not possible. However, thanks to the exponential dependence of the evaporation rate on mass, we may run simulations at lower masses and compare with analytic computations. This is still not completely straightforward, as particles can become caught in long orbits with very few interactions. Finite Monte Carlo time means that these particles will artificially extend the time recorded during which no evaporation has taken place, and thus artificially suppress the recorded rate. 

The top panel of Fig. \ref{fig:evapvsm} illustrates the evaporation rate recorded in this way for two dark matter-nucleon cross sections, $\sigma_\mathrm{SD} = 10^{-37}$ cm$^{2}$ (orange), and $\sigma_\mathrm{SD} = 10^{-36}$ cm$^{2}$ (blue). The bottom panel of Fig. \ref{fig:evapvsm} shows the evaporation rate for a 1 GeV (pink) and 2 GeV (green) DM particle as a function of interaction cross section.  In these figures the data points are from our MC simulations. The error bars are simply $\sqrt{N_{\mathrm{evap}}}/t_{\mathrm{sim}}$, and therefore do not account for the suppression mentioned above. This manifests as lower individual data points, especially at low masses. 

The lines in Fig. \ref{fig:evapvsm} show the evaporation rates computed as in Refs. \cite{Busoni2017EvaporationSun,Garani:2017jcj}. For the $m_\chi = 2$ GeV case in the lower panel, we show three different calculations: the top green curve (dotted) uses the isothermal approach, which overestimates evaporation, the bottom green curve (dot-dashed) uses the LTE number density distribution, and the middle green curve (dashed) shows the interpolation presented in \cite{Busoni2017EvaporationSun}. We were not able to obtain a convergence that was significant enough to fully evaluate these approaches deep in the LTE regime, however, based on our simulations, we conclude that calculations of the evaporation rate from the literature are trustworthy.

\begin{table*} [h]
    \centering
    \begin{tabular}{|c|c|c|c|c|} \hline
        \textbf{Body}   & \textbf{Interaction}  & \textbf{Mass (GeV)}   & $A$   & $K_0$ \\\hline\hline
                        &                       & 3                     & 0.427(3)              & 0.480(2) \\
                        & spin-                 & 5                     & 0.418(5)              & 0.441(3) \\
                        & dependent             & 10                    & 0.412(11)             & 0.350(6) \\
        Sun             &                       & 20                    & 0.418(15)             & 0.271(12) \\\cline{2-5}
                        &                       & 3                     & 0.525(8)              & 0.271(2) \\
                        & spin-                 & 5                     & 0.492(12)             & 0.325(5) \\
                        & independent           & 10                    & 0.470(22)             & 0.369(11) \\
                        &                       & 20                    & 0.463(44)             & 0.411(11) \\\hline
                        &                       & 1                     & 0.453(5)              & 0.423(4) \\
                        & spin-                 & 3                     & 0.394(16)             & 0.429(13) \\
                        & dependent             & 5                     & 0.430(11)             & 0.419(25) \\
        brown dwarf     &                       & 8                     & 0.426(21)             & 0.355(39) \\\cline{2-5}
                        &                       & 1                     & 0.527(8)              & 0.373(3) \\
                        & spin-                 & 3                     & 0.434(15)             & 0.440(12) \\
                        & independent           & 5                     & 0.440(25)             & 0.471(19) \\
                        &                       & 8                     & 0.438(16)             & 0.482(36) \\\hline
                        &                       & 50                    & 0.472(9)              & 0.429(6) \\
        Earth           & spin-                 & 100                   & 0.465(14)             & 0.479(7) \\
                        & independent           & 150                   & 0.440(13)             & 0.453(9) \\
                        &                       & 200                   & 0.451(17)             & 0.459(12) \\\hline
    \end{tabular}
    \caption{The values of the Knudsen transition parameter $K_0$ and the pre-factor $A$ defined in Eq.~\eqref{eq:SPrescale} for the calibrated Spergel \& Press formalism, computed to fit our simulation data.}
    \label{tab:params}
\end{table*}

\section{Conclusions}
\label{sec:conclusion}
We have simulated the equilibrium distribution and heat transport of dark matter in a variety of astrophysical bodies, where particles obey self-consistent equations of motion in a realistic background potential, density and temperature profile. Our simulations in a Solar model, a brown dwarf, and an Earth profile show that the corrected Spergel \& Press (SP) formalism (Eq.~\ref{eq:SPrescale}) provides a good parametrization in all the regimes that we have tested, for both spin-dependent and spin-independent dark matter-nucleon interactions. Most studies of heat transport in stars and their observable consequences in literature have adopted the corrected LTE (GR) approach that we have shown to mismodel heat transport; we therefore hope that this work provides further impetus for the community to use a more phenomenologically-verified approach. The \texttt{cosmion} software developed for this work is publicly available. We anticipate that a future release will also include heat transport by DM-electron scattering. We have also investigated evaporation of DM from the Sun, finding that the predicted rates from the literature are consistent with our simulation results. We leave a more thorough investigation of capture and evaporation to future work.

\acknowledgements
HB acknowledges partial support from the STFC HEP Theory Consolidated grants ST/T000694/1 and ST/X000664/1 and thanks other members of the Cambridge Pheno Working Group for useful discussions. ACV is supported by the Arthur B. McDonald Canadian Astroparticle Physics Institute, NSERC and the province of Ontario, with equipment funded by the Canada Foundation for Innovation and the Province of Ontario, and housed at the Queen’s Centre for Advanced Computing.
Research at Perimeter Institute is supported by the Government of Canada through the Department of Innovation, Science, and Economic Development, and by the Province of Ontario. 

\bibliography{references}

\appendix

\onecolumngrid

\section{Treatment of particles leaving the star}
\label{app:kepler}
If at any point the RKF45 integrator notes that the DM particle has exited the star prior to reaching the next collision point, we may terminate the integration. In this case we first check if $v > v_{esc}$, in which case we count the event as an \textit{evaporation} event, and start over initialising a new particle. 
If instead the particle remains bound, we first determine the time it takes to reach the surface from the previous collision point. If $v_r$ is negative, the particle must first turn around before exiting. In this case we use ``ghost collisions'' to propagate the particle until it turns around. We simply repeat steps 2-3 of the random walk algorithm above, without changing the particle's velocity at each collision point, until $v_r$ becomes positive. Once this is true, the trajectory can then be integrated to the solar surface with $r$ as the dependent variable. This first step is necessary as the RKF45 algorithm requires a monotonic dependent variable, which is not the case if $r$ must first reverse course.   

Upon exiting the star with $v < v_{esc}$, the particle follows a Keplerian orbit until re-entering the surface. The semi-major axis $a$ and eccentricity $e$ of the particle's orbit can be determined via its position $\vec{x}_{exit}$ and velocity $\vec{v}_{exit}$ as it exits the star by
\begin{equation}
    a=\frac{GM_\star R_\star}{2GM_\star-|\vec{v}_{exit}|^2R_\star}
\end{equation}
\begin{equation}
    e=\sqrt{1-\frac{|\vec{h}|^2}{GM_\star a}}\;,
\end{equation}
where $\vec{h}=\vec{x}_{exit}\times\vec{v}_{exit}$. The angle $\theta$ between the point of the orbit's closest approach to the star's centre and the point at which the particle exits the star is equal to the angle at which the particle re-enters the star. This angle is computed from the semi-major axis and eccentricity of the particle's orbit by
\begin{equation}
    \theta=\pm\cos^{-1}\left(\frac{a-e^{2}a-r}{er}\right).
    \label{keplerangle}
\end{equation}
The particle remains in the plane in which its exit position and velocity vectors lie, so its re-entry position can be computed using the angle computed in Eq.~(\ref{keplerangle}) via
\begin{equation}
    \vec{x}_{enter}=\cos(2\pi-2\theta)\,\vec{x}_{exit} + \sin(2\pi-2\theta)\,\left[\vec{h}\times\vec{x}_{exit}\right].
\end{equation}
The re-entry velocity is determined from the re-entry position using conservation of energy:
\begin{equation}
    \vec{v}_{enter}=\cos(\phi)\;\vec{x}_{enter}+\sin(\phi)\;\frac{\vec{h}}{|\vec{h}|}\times\vec{x}_{enter}\,,
\end{equation}
where $\phi=\cos^{-1}\left(\frac{v_{exit,r}}{|\vec{v}_{exit}|}\right)$, and  $v_{exit,r}$ is the radial component of the exit velocity.

To determine the time that the particle spent outside of the star, we first compute the area $A_{tot}$ enclosed by the total elliptical orbit followed by the particle, as well as the total orbital period $P$, as follows:
\begin{equation}
    A_{tot}=\pi a^2\sqrt{1-e^2}
\end{equation}
\begin{equation}
    T=\sqrt{\frac{4\pi^2}{GM}a^3}\;.
\end{equation}
We then determine the fraction of the total enclosed area swept out by the particle while outside of the star, and use Kepler's third law of orbital motion to relate that fraction of the area to the fraction of the total orbital period of the particle outside of the star. The area swept out between the particle and the centre of the star during its Keplerian orbit is
\begin{equation}
    A=a^2(e^2-1)\left[\frac{e\sin(\theta)}{e\cos(\theta)+1}
    -\frac{2\tanh^{-1}\left(\frac{(e-1)\tan(\frac{\theta}{2})}{\sqrt{e^2-1}}\right)}{\sqrt{e^2-1}}\right],
\end{equation}
and the time $\Delta t$ spent outside of the star is then
\begin{equation}
    \Delta t=(1-A/A_{tot})\,P\,.
\end{equation}

\section{Quickstart Guide to \cosmion}
\label{app:quickstart}
The following are step-by-step instructions on how to use \cosmion:

\begin{enumerate}
  \item First, ensure you have a Fortran compiler installed on your system. This guide assumes the GNU Fortran compiler (gfortran). Fortran compilers can be specified in the Makefile. 
  \item Compile the shared library and the program using the provided Makefile with the commands:
  \begin{lstlisting}[language=bash]
  make csharedlib.so
  make cosmion.x
  \end{lstlisting}
  This creates a shared library named \texttt{csharedlib.so} and an executable named \texttt{cosmion.x}.

  \item Run the program with the command:
  \begin{lstlisting}[language=bash]
  ./cosmion.x <massin> <sigmain> <Nstepsin> <FileNameIn> <SpinDepIn> <vdepin>
  \end{lstlisting}
  where:
  \begin{itemize}
    \item \texttt{<massin>}: The mass input, in GeV.
    \item \texttt{<sigmain>}: Cross-section in cm$^2$.
    \item \texttt{<Nstepsin>}: The number of requested collisions.
    \item \texttt{<FileNameIn>}: The output filename.
    \item \texttt{<SpinDepIn>}: The spin-dependence input, either ``nucleonSI" for spin-independent or ``nucleonSD" for spin-dependent.
    \item \texttt{<vqDepIn>}: velocity or momentum-dependence: ``const" for constant interactions. Other options are v2, q2, v4, q4, vm2, qm2, respectively.
  \end{itemize}

  \item The program will print its progress to the console and write its output to the file specified by \texttt{<FileNameIn>}.

  \item The output file contains the simulation results. Each line in the file corresponds to a single step of the simulation and contains: 
    \begin{itemize}
      \item Positions (x, y, z) in cm
      \item Velocities (v\_x, v\_y, v\_z) in cm/s
      \item Time in s
      \item The flag that indicates the current state of the particle
      \item The weight of the current position      
    \end{itemize}
\end{enumerate}

For example:
  \begin{lstlisting}[language=ba    sh]
./cosmion.x 10 1.0d-36 1000000 "output.txt" nucleonSI const
  \end{lstlisting}
will launch a simulation of one million collisions for a 10 GeV dark matter particle, with a $10^{-36}$ cm$^{2}$ spin-independent DM-nucleon cross section, to be written to the output file \texttt{output.txt}. 
\\

The \texttt{cosmion.f90} program file also contains a number of logical flags that can affect the simulation behaviour:

\begin{tabularx}{\textwidth}{|>{\hsize=.3\hsize}X|>{\hsize=1.7\hsize}X|}
\hline
\textbf{Flag} & \textbf{Description} \\
\hline
debug\_flag & Used for debugging. If set to \texttt{.true.}, the program outputs additional information for troubleshooting. \\
\hline
spinDep & Determines the spin-dependence of the collisions. If set to \texttt{.true.}, the collisions are spin-dependent and only hydrogen is considered. If set to \texttt{.false.}, the number of isotopes to use is determined by the \texttt{species\_precision} value. This flag  is set by command-line input.\\
\hline
anTemp & If set to \texttt{.true.}, the star's temperature is determined using analytic functions defined in \texttt{star.f90}. If set to \texttt{.false.}, the temperature is interpolated from a stellar model. \\
\hline
anDens & If set to \texttt{.true.}, the star's density is determined using analytic functions defined in \texttt{star.f90}. If set to \texttt{.false.} (default), the density is interpolated from a stellar model. \\
\hline
anPot & If set to \texttt{.true.}, the potential is treated as a simple harmonic oscillator (SHO) and analytic trajectory expressions are used for \texttt{x} and \texttt{v}. If set to \texttt{.false.} (default), the potential is determined from a stellar model. 
\\
\hline
SHO\_debug & If set to \texttt{.true.}, the tabulated $\phi(r)$ is overridden to provide a simple harmonic oscillator (SHO) potential, but trajectories are still solved numerically. This flag is only used for testing that the integrator is working properly (for comparison with analytic trajectories), and has no effect if \texttt{anPot} is already \texttt{.true.}. \\
\hline
outside\_flag & This flag is used to keep track of the current state of the particle. The possible values are 0 (inside the star), 1 (leaving the star), 2 (escaped). A value of -1 is used internally and should not be seen in the outputs. \\
\hline
\end{tabularx}

\section{Collision rates per nuclear species}
\label{app:collisionrate}
\begin{figure}[h]
    \centering
    \includegraphics[width=0.9\columnwidth]{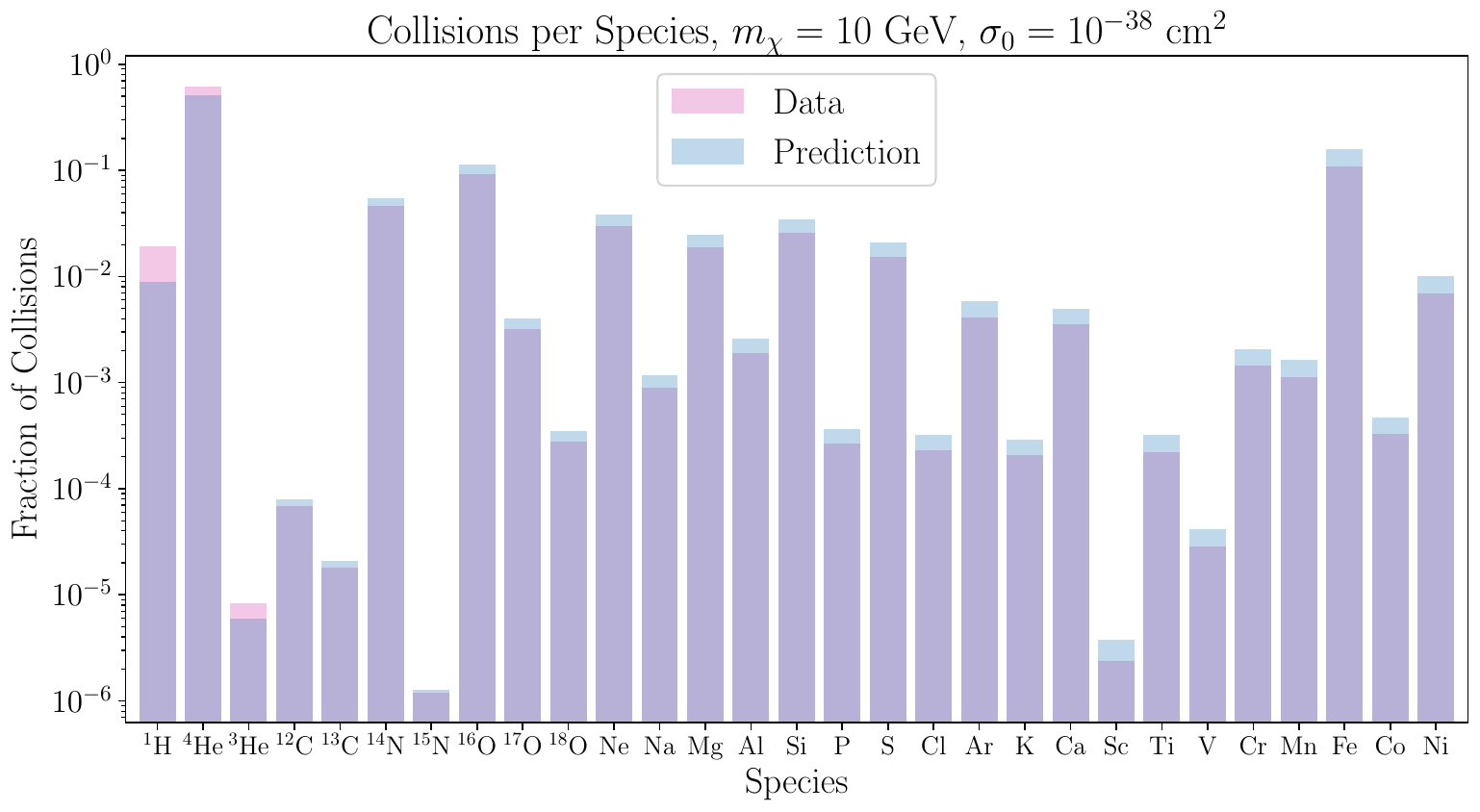}
    \caption{The fraction of collisions of a $m_\chi=10$ GeV DM particle with each of the 29 most abundant nuclear species in the Sun. The particle has a constant, spin-independent interaction cross-section of $\sigma_0=10^{-38}$ cm$^2$. The pink bars represent the fraction of collisions from a simulation in a realistic gravitational potential containing $N=10^8$ collisions. The blue bars are the fraction of collisions predicted by the code prior to the simulation, which allows for species precision selection.}
    \label{fig:speciesBar}
\end{figure}

\begin{figure}[h]
    \centering
    \includegraphics[width=0.9\columnwidth]{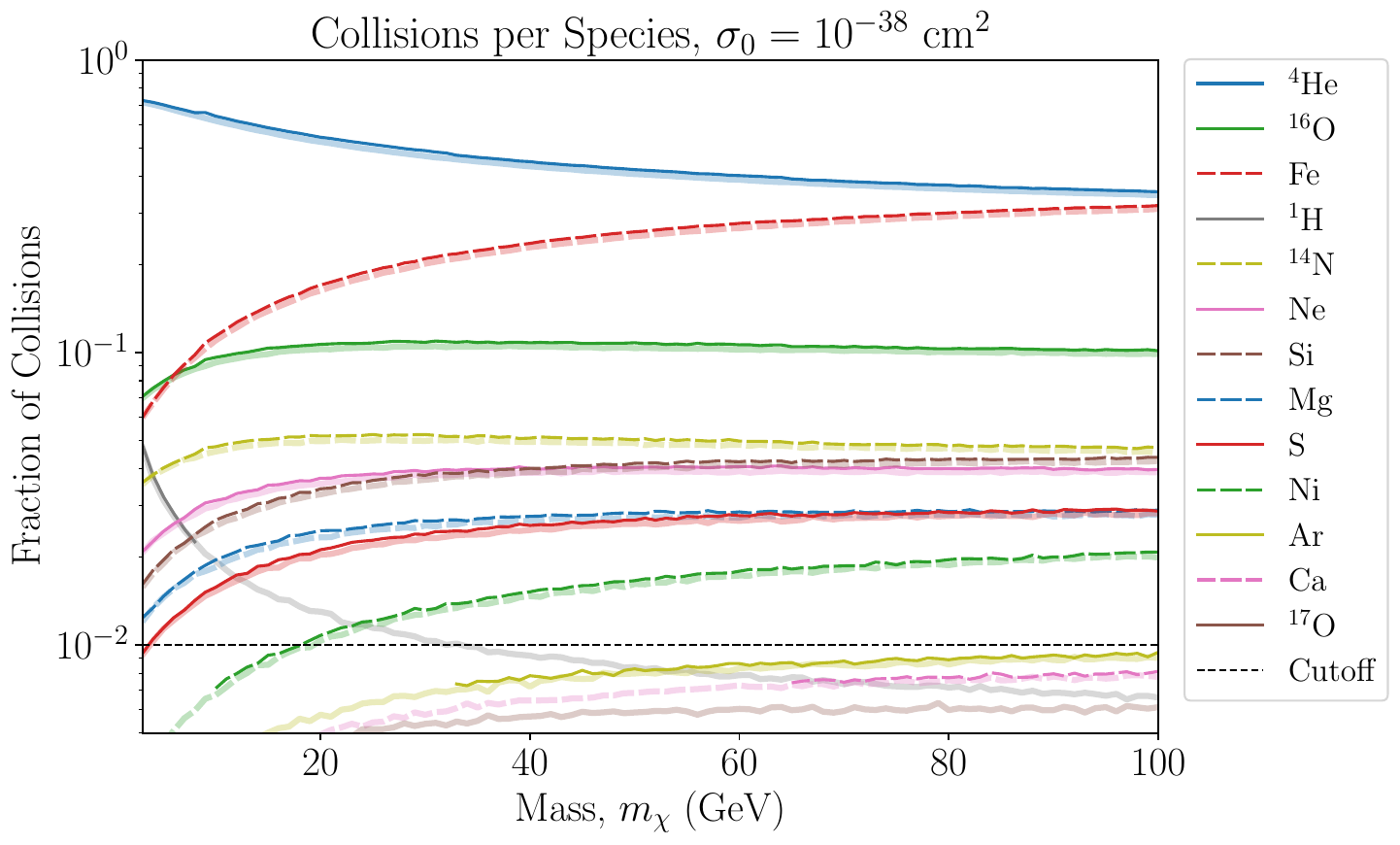}
    \caption{The fraction of collisions of the DM particle with the most impactful of the 29 most abundant nuclear species in the Sun, as a function of the dark matter particle mass $m_\chi$. A simulation containing $N=10^6$ collisions was carried out at each integer particle mass $m_\chi$ in GeV, each with spin-independent interaction cross-section of $\sigma_0=10^{-38}$ cm$^2$. The thick pale lines result from simulations considering all 29 nuclear species. The thin dark lines result from simulations with a species selection precision of $P=10^{-2}$, whose intended cutoff precision is represented by the horizontal short-dashed line.}
    \label{fig:speciesMasses}
\end{figure}

In this appendix, we show the collision rate with each species in the Sun. The \texttt{cosmion} code can predict these rates, and truncate the number of species based on a desired precision. Fig. \ref{fig:speciesBar} shows the collision rate with each species for a 10 GeV dark matter particle, as predicted by this algorithm (blue bars) and as obtained in a simulation with $10^8$ collisions. Fig. \ref{fig:speciesMasses} displays the rates as a function of dark matter mass. We illustrate how, if the user had selected a precision of $10^{-2}$, species falling below the dotted line would be omitted from the simulation for speed.

\end{document}